\documentclass[nofootinbib,aip,graphicx]{revtex4-1}
\usepackage{epsfig}
\usepackage{amssymb}
\usepackage{amsmath}
\usepackage{epstopdf}
\usepackage{subfigure}
\usepackage{subfloat}
\usepackage{lineno}
\usepackage{float}
\usepackage{color}
\usepackage{units}
\usepackage{enumerate}
\usepackage{multirow}
\usepackage{rotating}
\usepackage{array}
\usepackage[version=4]{mhchem}
\usepackage{chemist}
\usepackage{tikz,chemfig}
\usepackage{blkarray}
\usepackage{subfigure}
\usepackage{gensymb}
\usepackage{amsmath,amssymb,mathrsfs,bm,amsthm}
\usepackage{array}
\usepackage{mathtools}
\usepackage[normalem]{ulem}
\usepackage{dcolumn}
\usepackage[caption=false]{subfig}

\usepackage{mathrsfs,amsmath,amssymb,mathtools} 
\draft % marks overfull lines with a black rule on the right

\begin{document}

\preprint{}

\title{Lindblad parameters from high resolution spectroscopy to describe collision induced ro-vibrational decoherence in the gas phase -- Application to acetylene} %Title of paper
%\title{Collision induced decoherence in the gas phase, Lindblad parameters from high resolution spectroscopy -- Application to acetylene}


\author{Antoine Aerts}
\author{Jean Vander Auwera}
\author{Nathalie Vaeck}
\email[Email: ]{nvaeck@ulb.ac.be}
%\thanks{}
\affiliation{Universit\'e Libre de Bruxelles; Spectroscopy, Quantum Chemistry and Atmospheric Remote Sensing (SQUARES); 50 avenue F. Roosevelt, C.P. 160/09, B-1050 Brussels, Belgium}


\date{\today}
\textit{This article may be downloaded for personal use only. Any other use requires prior permission of the author and AIP Publishing. This article appeared in }(J. Chem. Phys. 154, 144308 (2021)) \textit{and may be found at }(\url{https://doi.org/10.1063/5.0045275}).
\begin{abstract}

Within the framework of the Lindblad master equation, we propose a general methodology to describe the effects of the environment on a system in dilute gas phase. The phenomenological parameters characterizing the transitions between rovibrational states of the system induced by collisions can be extracted from experimental transition kinetic constants, relying on Energy Gap fitting laws. As the availability of this kind of experimental data can be limited, the present work relied on experimental line broadening coefficients, however still using Energy Gap fitting laws. The 3 $\mu$m infrared spectral range of acetylene was chosen to illustrate the proposed approach. The method shows fair agreement with available experimental data while being computationally inexpensive. The results are discussed in the context of state laser quantum control.

\end{abstract}


\maketitle %\maketitle must follow title, authors, abstract and \pacs
    
%-------------------------------------------------------

% \section{}
%\label{}
% \subsection{}
% \subsubsection{}

% If in two-column mode, this environment will change to single-column format so that long equations can be displayed. 
%\begin{widetext}
%$$\mbox{put long equation here}$$
%\end{widetext}

% Figures should be put into the text as floats. 
% Use the graphics or graphicx packages (distributed with LaTeX2e).
% Put the label that you will use with \ref{} command in the braces of the \label{} command.
%
% \begin{figure}
% \end{figure}

% \begin{table}
% \end{tabular}
% \end{table}
    
%-------------------------------------------------------

With the increasing availability of computational power and promises of the quantum technologies, the impact of decoherence processes caused by the interactions with the environment is receiving more and more attention in studies of the dynamics of realistic open systems.\cite{jpcm_28_213001} It is worth to mention the importance of decoherence in the context of molecular rotation\cite{rmp_91_035005} control, which is particularly relevant in this case, but also in biological systems,\cite{jppac_190_372,NJP_16_045007} molecular dynamics,\cite{pra_78_063408,jcp_148_134304} vibration of adsorbed molecules\cite{jcp_128_194709,prb_100_245431} or chemical reactions.\cite{jpca_116_11273,jpb_50_082001} In the context of quantum control, decoherence caused by the interaction of the system of interest with its environment is a major issue for applications. Decoherence can act against the control performance or induce a loss of ``quantum" property (loss of coherence). Although, numerous theoretical control studies were mainly conducted without assessment of the environment-induced processes, the attention is now to include dissipative environments.\cite{prl_95_113001,jcp_124_154105,jcp_142_134304}

Within the scope of rovibration dynamics and its eventual control, the primary source of  perturbation, inherent to the system, arises from intramolecular vibrational redistribution (IVR) and was discussed before.\cite{mp_113_4000} It is included in the simulation dynamics by design.\cite{mp_113_4000,mp_116_2213} All environmental effects acting on the molecule are expected to cause decoherence that, in the gas phase, is well pictured by the alteration of the shape of the spectral lines.\cite{aj_2_251} It can easily be shown that collisions are the main source of shape alteration at atmospheric pressure, giving rise in the time domain to population transfers from one rotational level to another. 

Rotational decoherence is particularly important in the understanding of dissipation occurring in gases.\cite{NC_10_5780,jcp_136_184302} Its effects can be studied by a variety of experimental techniques including infrared-ultraviolet double resonance (IRUVDR)\cite{jms_84_272}, molecular centrifuges more suitable for studies of high angular momenta\cite{prl_82_3420,prl_85_542} as well as echo spectroscopy.\cite{NC_10_5780} The rotational relaxation can also be computed \textit{ab initio},\cite{jcp_58_5422} however requiring a significant computational effort to obtain the intermolecular potential and to calculate the $S$-matrix. An alternative method involves the study of lineshapes in the spectral domain. Collision-induced decays or transfers of rotational populations indeed condition the shape of spectral lines, as modeled by the relaxation matrix.\cite{levy_1992,hartmann_2008} The elements of this matrix can be measured experimentally or calculated. In particular, the real part of the relaxation matrix can be constructed in a rather simple way using Energy Gap fitting laws.\cite{hartmann_2008} Energy Gap fitting laws arise from the intuitive picture that rotational relaxation should drop with the change of rotational quantum number $J$, or equivalently with the energy difference, induced by the collisions. This intuitive picture was well reproduced in experimental studies since its first mention,\cite{jcp_56_1563} and has spawned the development of numerous energy based scaling laws,\cite{brunner_1982} which for example proved to be successful in the description of line mixing effects in collisionally-broadened CO$_2$ infrared branches.\cite{jqsrt_36_521,jcp_84_1149,jcp_86_5722}

The aim of the present work is to couple high resolution spectroscopy with quantum dynamics in the context of collision dynamics in the gas phase. High resolution spectroscopic data on acetylene (effectively its main isotopologue $^{12}$C$_2$H$_2$) were used to illustrate the purpose. The time evolution induced by collisions of the population of a rovibrationally excited level of acetylene in a dilute gas phase is calculated using the Lindblad master equation, the required (Lindblad) parameters being obtained from either pump-probe experiments or collisional line broadening coefficients, relying on Energy Gap fitting laws. Approaches used in this work are well established in their respective communities; this article tries to present them in an unified manner. The present work is part of our effort to better understand the dynamics and the control of the vibrational population of acetylene to possibly access the vinylidene isomer. It was shown that a specific vibrational mode of acetylene (\ce{C2H2}) in dilute gas phase can be populated using a single shaped laser pulse.\cite{mp_113_4000} In other words, a state well isolated from others with respect to IVR could be specifically populated via the control of the population of vibrational levels of \ce{C2H2}. This study was later\cite{mp_116_2213} extended by inclusion of the rotational structure in the description of the controlled system state, which proved to be essential for the determination of the control pulse shape. The purpose of this work is to propose a methodology to allow taking into account the effect of the environment on the dynamics given the collision-induced rotational state-to-state transfers.

In the next sections, the vibration-rotation structure of the acetylene molecule is briefly introduced followed by the Lindblad master equation that describes the quantum dynamics of the system, including the effects of its environment. The determination of the Lindblad parameters from pump probe experiments and collisional line broadening coefficients is then detailed. This article concludes with some perspectives. %We also comment on the construction of this relaxation matrix from fitting of broadening coefficients and on the simulation of spectra.

%-------------------------------------------------------

\section{The vibration-rotation structure of acetylene}

The vibrational motion of the main isotopologue of the acetylene molecule, $^{12}$C$_2$H$_2$, can be described in terms of the 5 modes of vibration presented in Table \ref{table:normalmodes}. The zero-order vibrational levels associated with the excitation of these modes are identified by $|v_1 v_2 v_3 v_4^{\ell_4} v_5^{\ell_5}\rangle$, where $v_i$ is the vibrational quantum number associated with the mode of vibration $\nu_i$ ($i=1\dots5$, see Table \ref{table:normalmodes}) and $\ell_4$ and $\ell_5$ are the quantum numbers associated with the vibrational angular momenta generated by the excitation of the doubly degenerate bending modes $\nu_4$ and $\nu_5$, respectively. These vibrational levels interact mostly through anharmonic resonances, leading to the formation of so-called polyads.\cite{jpcrd_45_023103}

The present work focuses on the polyad of vibrational levels of the ground electronic state of $^{12}$C$_2$H$_2$ that gives rise to the strong 3 $\mu$m infrared absorption of this molecule when accessed from the vibrational ground  state. This polyad involves 3 zero-order levels, namely $|0010^00^0\rangle$, $|0101^11^{-1}\rangle$ and $|0101^1 1^1\rangle$. Allowed one-photon transitions from the ground state only involve the $|0010^{0}0^{0}\rangle$ level. It is therefore identified as a ``bright'' level, while the other two are identified as ``dark'' levels. Anharmonic resonances strongly mix the $|0010^{0}0^{0}\rangle$ and $|0101^{1}1^{-1}\rangle$ levels, while the $|0101^{1}1^{1}\rangle$ level essentially keeps its ``dark'' character.\cite{mp_113_4000,jms_157_337} As this latter level was shown to not being involved in the collisional processes considered here,\cite{jcp_98_8572,jcsft_94_3219} it will not be considered any further. The eigenstates resulting from the mixing of the former two levels are identified from now on as $\Gamma_1$ and $\Gamma_2$, $\Gamma_2$ having the highest energy of the two. Transitions from the ground state to $\Gamma_1$ and $\Gamma_2$ result into two strong bands observed near 3 $\mu$m, which are called experimentally the $\nu_2 + (\nu_4 + \nu_5)^0_+$ and $\nu_3$ bands, respectively. The exponent 0 appearing for the former band is the value of $k = \ell_4 + \ell_5$ and the ``+'' sign refers to the symmetry of the vibrational wavefunction with respect to the infinity of planes containing the molecule.\cite{bunker_1979} 

In section \ref{section:linebroad}, self broadening coefficients of vibration-rotation lines of the $\nu_3$ band are needed. However, they have never been directly measured for this band, most probably because of its strength. Fortunately, the vibrational dependence of self broadening coefficients is generally small and has never clearly been evidenced for acetylene.\cite{jqsrt_75_397} In the frame of the present work, we therefore relied on the extensive set of self broadening coefficients measured for the $3\nu_5^1$, $(2\nu_4+\nu_5)^{1}$I and $(2\nu_4+\nu_5)^{1}$II bands observed in the 5 $\mu$m region ($1860-2180$ cm$^{-1}$).\cite{jqsrt_75_397} The Roman numerals I and II discriminate the two $k=1$ vibrational levels with the same symmetry that arise from the simultaneous excitation of 2 quanta in $\nu_4$ and one quantum in $\nu_5$, the Roman numeral I identifying the level characterized by the highest energy.\cite{jmsp_44_145,jmsp_44_165}

Acetylene is a linear molecule in its ground electronic state. Therefore, the description of its rotational motion mainly involves the quantum number $J$ associated with the angular momentum of the molecule.

\begin{table}
\begin{tabular}{|c|c|c|}
\hline
% \setatomsep{1.8em}
Normal mode&\chemfig{H-C~C-H} &Harmonic frequency / cm$^{-1}$ \\
\hline
$\nu_1$  &
\includegraphics{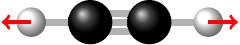}  &3389.12\\
\hline
$\nu_2$  &
 \includegraphics{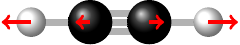} &1950.11\\
\hline
$\nu_3$  &
\includegraphics{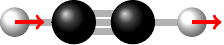}& 3310.02\\
\hline
$\nu_4$  &
\includegraphics{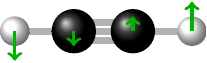}  & 604.47\\
\hline
$\nu_5$  &
\includegraphics{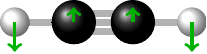}  & 728.27\\
\hline
\end{tabular}\par
\caption{Normal modes of vibration of acetylene $^{12}$C$_2$H$_2$ in its ground electronic state and their harmonic frequencies.\cite{mp_105_2217} The first three are non degenerate stretching modes and the last two are doubly degenerate bending modes.}
\label{table:normalmodes}
\end{table}
    
%-------------------------------------------------------

\section{Lindblad equation}

The evolution of a system coupled to the environment in the Born and Markov approximation can be described by the Lindblad master equation:\cite{cmp_48_119,jmp_17_821}
\begin{align}
\frac{d \hat{\rho}(t)}{d t}=&-\frac{i}{\hbar}\left[ \hat{H}(t),\hat{\rho}(t)\right] \label{eq:lindblad}\\ 
\nonumber
&+ \sum_{ij}\left(\hat{L}_{ij}\hat{\rho}(t)\hat{L}_{ij}^\dagger -\frac{1}{2} \left[\hat{\rho}(t),\hat{L}_{ij}^\dagger \hat{L}_{ij}\right]_+\right) 
\end{align}
where 
\begin{equation}
\hat{H}(t)= \hat{H}_S-\hat{\mu}^e\epsilon(t)
\label{Ht}
\end{equation}
is the total Hamiltonian in the absence of interaction with the environment and $\hat{H}_S$ the Hamiltonian of the system. If a laser field is applied, $\hat{\mu}^e$ is the projection of the electric dipole moment of the system along the direction of the polarization of the field and $\epsilon(t)$ is the time-dependent amplitude of the laser pulse. $[A,B]_+$ is the anti-commutator of arguments $A$ and $B$, $\hat{\rho}(t)$ is the time-dependent density operator and $\hat{L}_{ij}$ are operators representing the transitions induced by the environment between states $i$ and $j$ of the system.
A phenomenological representation of the transition operator was used:
\begin{equation}
\label{eq:Lindblad_frequences}
\hat{L}_{ij}=\sqrt{\theta_{ij}}|i\rangle \langle j|.
\end{equation}
With this form, decoherence in the system is characterised by the Lindblad parameters $\theta_{ij}$, which are the transition rates induced by non-radiative processes (called transition rates in the text, the use of $\theta$ is a deliberate choice; $\gamma$ is generally used in the literature) between eigenstates $|i\rangle$ and $|j\rangle$ of the system Hamiltonian. Comments on the implementation of the Lindblad master equation are given in the Supplemental Material. The determination of the Lindblad parameters $\theta_{ij}$ from experimental data published in the literature is discussed in the next two sections.
    
%-------------------------------------------------------

\section{Lindblad parameters from pump-probe experiments}
\label{Previous}

Frost\cite{jcp_98_8572} and Henton \textit{et al.}\cite{jcsft_94_3219} studied experimentally the dynamical effects of collisions in the $\Gamma_1$ and $\Gamma_2$ interacting states using infrared ultraviolet double resonance (IRUVDR). An infrared ``pump'' laser of fixed wavelength was used to populate a specific rotational level of the eigenstate $\Gamma_1$ or $\Gamma_2$ and an ultraviolet ``probe'' laser induced a transition from the excited rovibrational level to the first excited electronic state. Collision-induced transitions within the eigenstates $\Gamma_1$ and $\Gamma_2$ were monitored from the observed evolution of the laser induced fluorescence with the probe laser wavelength. The corresponding state-to-state rate constants for rotational relaxation were determined. Frost\cite{jcp_98_8572} initiated the study for self-relaxation (C$_2$H$_2$ in an environment of C$_2$H$_2$), while Henton \textit{et al.}\cite{jcsft_94_3219} refined his measurements and extended this research to the effects of foreign gases (Ar, He and H$_2$) on the dynamics.

\begin{figure}[htbp]
\centering
\includegraphics[width=\linewidth]{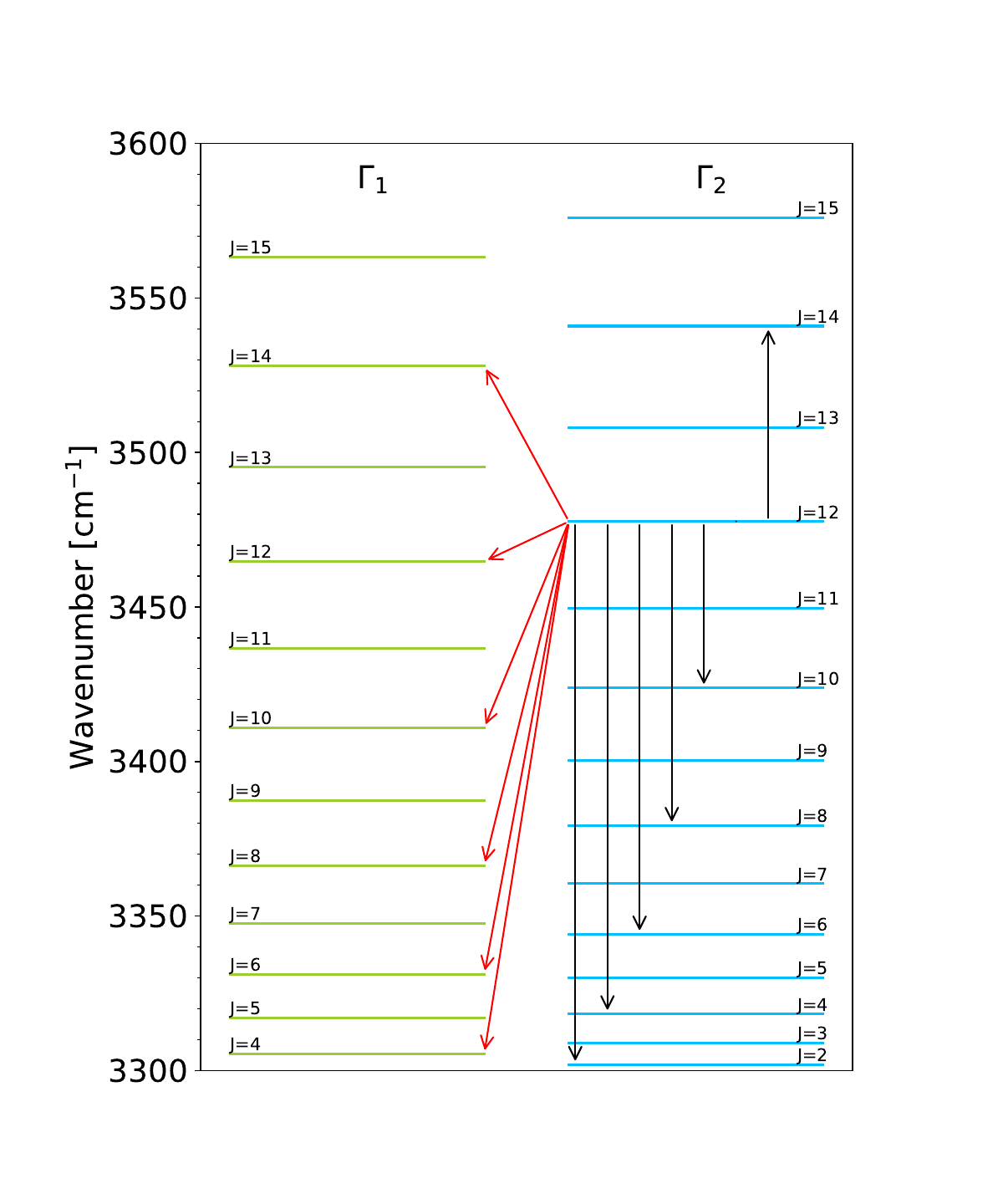} 
\caption{Collision-induced state-to-state transfers from the $\Gamma_2, J=12$ level probed by the IRUVDR experiments. Relaxation processes R1 and R3 are explicitly depicted in red and black respectively.}
\label{fig:statetostate}
\end{figure}

Frost\cite{jcp_98_8572} and Henton \textit{et al.}\cite{jcsft_94_3219} considered four relaxation processes caused by collisions. They are described as follows :
\begin{alignat}{5}  
&\ce{C_2H_2} [ \Gamma_{2}\phantom{|};J_i ] & \ce{+ M} & \ce{-> }& \ce{C_2H_2} [ \Gamma_{1}\phantom{|};J_j ] & \ce{+M} +\Delta E \label{equ:react_intradyadplus}\tag{R1}\\
&\ce{C_2H_2} [ \Gamma_{1}\phantom{|};J_i ] & \ce{+ M} & \ce{-> }& \ce{C_2H_2} [ \Gamma_{2}\phantom{|};J_j ] & \ce{+M} +\Delta E \label{equ:react_intradyadmoins}\tag{R2}\\
&\ce{ C_2H_2} [ \Gamma_{2}\phantom{|};J_i ] & \ce{+ M} & \ce{-> }& \ce{C_2H_2} [ \Gamma_{2}\phantom{|};J_j ] & \ce{+M} +\Delta E \label{equ:react_rr2}\tag{R3}\\
&\ce{ C_2H_2} [ \Gamma_{1}\phantom{|};J_i ] & \ce{+ M} &\ce{ ->} & \ce{C_2H_2} [ \Gamma_{1}\phantom{|};J_j ] & \ce{+M} +\Delta E \label{equ:react_rr1}\tag{R4}
\end{alignat} 
where M is the collision partner (\textit{i.e.} other $\ce{C_2H_2}$ molecules in the present case), $J_i$ and $J_j$ are respectively the initial and final rotational levels and $\Delta E$ is the energy released/absorbed during the collision. It is worth noting that these processes are subject to the parity selection rule\cite{oka_1974} that forbids the change of the total wavefunction parity,\cite{jce_59_17} \textit{i.e.} ortho $\leftrightarrow$ para transitions are not observed.\cite{jcp_98_8572,jcsft_94_3219} Ortho and para label the levels with the highest and lowest nuclear spin statistical weights, respectively. In \ce{^12C2H2}, ortho (\textit{resp.} para) labels levels with odd (\textit{resp.} even) $J$. Consequently, non-radiative transitions involving even $\Delta J = |J_j-J_i|$ are solely observed in experiments. The eigenstate level energies and the rotational state-to-state transitions probed by the experiments are illustrated in Fig. \ref{fig:statetostate}.

Frost \cite{jcp_98_8572} and later Henton \textit{et al.} \cite{jcsft_94_3219} determined the transition kinetic constants $k(j\leftarrow i)$ of processes R1 to R4 using the IRUVDR technique. They fitted their measurements to Exponential Gap Laws (EGL) of the form:
\begin{equation}
k(j\leftarrow i|K_0,\eta)=K_0\phantom{-}\exp\left(\frac{-\eta|E_{j}-E_{i}|}{k_BT}\right) \phantom{ - - } \text{with } E_j>E_i
\label{EGL}
\end{equation}
where $K_0$ (in units of frequency per pressure as $k$) and $\eta$ (dimensionless) are fitted parameters and $E_i$ is the energy of level $i$. The downwards transition kinetic constants are deduced from Eq. \ref{EGL} and the detailed balance (to ensure conservation of total population):
\begin{equation}
k(i\leftarrow j)=\frac{\rho_i}{\rho_j}k(j\leftarrow i)=\frac{2J_i+1}{2J_j+1}\exp\left(\frac{E_j-E_i}{k_BT}\right)k(j\leftarrow i)
\label{equ:balance}
\end{equation}
with $\rho_i$ the relative population of level $i$, proportional to $(2J_i+1)\exp\left(-E_i/k_BT\right)$. 
Their reported parameters and uncertainties are reproduced in Table \ref{table:EGL_jcsft_94_3219}, together with the identification of the level populated by the pump laser. Table \ref{table:EGL_jcsft_94_3219} shows that the $\eta$ parameters of the 5 different processes have different fitted values although the reported uncertainties are large. Within the uncertainties, all $\eta$ parameter ranges overlap. On the other hand, comparisons of the magnitude of the $K_0$ parameters show that transitions within the same vibrational eigenstate (R3 and R4) are an order of magnitude faster than transitions involving a change of eigenstate (R1 and R2) and that transitions with $\Delta J =\pm 2$ in the $\Gamma_1$ eigenstate are greatly favoured. In this later case, the $\eta$ parameter was fixed by Henton \textit{et al.}\cite{jcsft_94_3219} to the value reported for the $\Delta J \ge 4$ transitions, given the expected strong correlation of the parameters.\cite{jcp_106_3592}

\begin{table}[h!]
\begin{center}
\begin{tabular}{lD{.}{.}{5.7}D{.}{.}{5.8}l}
\hline
\hline
Process & \multicolumn{1}{c}{$K_0$} & \multicolumn{1}{c}{$\eta$} & Initial \\
\hline
\ref{equ:react_intradyadplus}         & 0.005(1)  & 1.7(7)   & $\Gamma_2$, $J=12$ \\
\ref{equ:react_intradyadmoins}        & 0.0036(3) & 0.91(31) & $\Gamma_1$, $J=10$ \\
\ref{equ:react_rr2}                   & 0.032(3)  & 1.92(17) & $\Gamma_2$, $J=12$ \\
\ref{equ:react_rr1}, $|\Delta J|=2$   & 0.045(7)  & 1.71     & $\Gamma_1$, $J=10$ \\
\ref{equ:react_rr1}, $|\Delta J|\ge4$ & 0.019(3)  & 1.71(27) & $\Gamma_1$, $J=10$ \\
\hline
\hline
\end{tabular}
\caption{Experimental parameters ($K_0$ in cm$^{-1}$atm$^{-1}$ and $\eta$ is dimensionless) of the EGL law (Eq. \ref{EGL}) from Frost \cite{jcp_98_8572} and Henton \textit{et al.} \cite{jcsft_94_3219} used in this work to calculate the kinetic constants of processes \ref{equ:react_intradyadplus} to \ref{equ:react_rr1}. Numbers between parentheses are the standard deviations in the units of the last digit quoted. The last column identifies the level initially populated by the pump laser.}
\label{table:EGL_jcsft_94_3219}
\end{center}
\end{table}

To express the Lindblad parameters $\theta_{ij}$ (see Eq. \ref{eq:Lindblad_frequences}) from this data set, we assumed that they are related to the transition kinetic constants $k(j\leftarrow i)$ by: 
\begin{equation}
\theta_{ij}=P\phantom{|}k(j\leftarrow i)
\label{equ:EGL}
\end{equation}
The dependence with pressure is assumed to follow the same linear dependence as that associated with line broadening, the coefficients of which being expressed as a sum of $k(j \leftarrow i)$ at sub-atmospheric pressures.\cite{hartmann_2008} A correction to the $\theta$ parameters may however be needed at higher pressure,\cite{jcp_45_1649} or even a different approach,\cite{jcp_106_8299} because the IRUVDR measurements could not be extrapolated to that regime and equations given in the present work would not hold. The current assumption is based on the relation of transition rates with the broadening coefficients (see Eq. \ref{relax1} in the next section) and the common linear dependency of the broadening coefficient  with respect to pressure in high resolution spectroscopy where the reference value is taken at $P=1$ atm.

To illustrate the collision-induced dynamics, the evolution of the population of selected levels of the polyad of interest at $P=1$ atm and $T=298$ K (Eqs \ref{EGL} and \ref{equ:EGL}) is shown, as described by the Lindblad master equation (Eq. \ref{eq:lindblad}) without interaction with an external laser field. The transition rates $\theta_{ij}$ are obtained from Eqs. \ref{equ:EGL} and \ref{EGL}, and the experimental EGL parameters given in Table \ref{table:EGL_jcsft_94_3219}. The $H_S$ hamiltonian (Eq. \ref{Ht}) of acetylene itself is described in the state space by a global effective hamiltonian,\cite{jpcrd_45_023103} built from the extensive set of experimental high-resolution spectroscopic studies carried out in the ground electronic state of the molecule. More details can be found in our previous work,\cite{mp_113_4000,mp_116_2213} and a thorough and pedagogical introduction to the strategy applied to build such global hamiltonians was given by Herman.\cite{herman_handbook} Working in the state space connects the relevant quantities obtained from high resolution spectroscopy, \textit{i.e.} eigenstate energies from line positions and transition dipole moment operators from line intensities, to the system dynamics. The state basis used for the simulation includes levels from $J$=0 to $J$=100. The initial population lies in the $\Gamma_2, J=12$ rovibrational level as an illustration of the dynamics probed by the experiments of Frost.\cite{jcp_98_8572}
\begin{figure}[htbp]
\centering
\includegraphics[width=\linewidth]{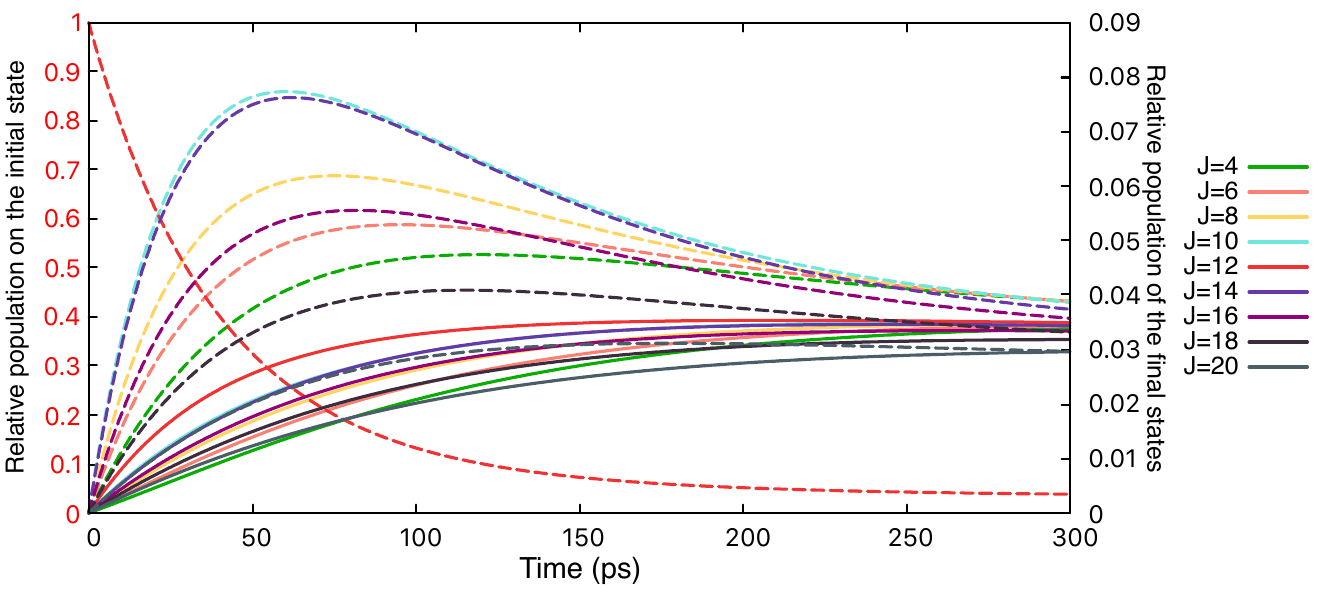} 
\caption{Collision induced dynamics in $\Gamma_1$ (\textbf{continuous}, scale on the right axis) and in $\Gamma_2$ (\textbf{dashed}) from the singly populated $\Gamma_2, J=12\rangle$ state (shown in red on the left axis) under $P=1$ atm and $T=298$ K. State-to-state transition rates are calculated from Eqs. \ref{equ:EGL} and \ref{EGL}, and the parameters reported by Frost \cite{jcp_98_8572} and Henton \textit{et al.}\cite{jcsft_94_3219} (reproduced in Table \ref{table:EGL_jcsft_94_3219}).}
\label{fig:PCL_collisions}
\end{figure}
The dynamics is solely driven by population transfers between rotational levels induced by inelastic collisions as illustrated in Fig. \ref{fig:statetostate}, \textit{i.e} transfers from/to another polyad or vibrational level except $\Gamma_1$ and $\Gamma_2$ are neglected as are parity-violating transitions that happen on a longer timescale. Fig. \ref{fig:PCL_collisions} shows the relative population ($\frac{\rho_{\Gamma_i, J}(t)}{ \rho_{\Gamma_2, J=12}(t_0) }$) of  the levels $\Gamma_1$, $\Gamma_2$ and $J=4$ to 20 from initial time $t=t_0$. The simulation shows a rapid population transfer from the initially populated state to other states (almost complete within 100 ps), with a larger propensity for transfers within the same vibrational eigenstate. The $\Gamma_2, J=12$ curve in Fig. \ref{fig:PCL_collisions} is a good illustration of the first-order kinetics introduced by the Lindblad equation. For one eigenstate system dynamics, \textit{i.e.} a system with a purely diagonal system hamiltonian ($\hat{H}_S$) by definition,   driven solely by interaction with its environment $\left(-\frac{i}{\hbar}\left[ \hat{H}(t),\hat{\rho}(t)\right]=0\right)$, Eq. \ref{eq:lindblad} simplifies to a sum of products of scalars and density matrices when the phenomenological representation of the transition operators is used (see Eqs. S3 and S4 in the Supplemental Material).

All expected characteristics of the dynamics are reproduced in the simulation. The downwards transitions ($\Delta J < 0$) are favoured and reach an higher maximum population than their upwards counterparts ($\Delta J > 0$) given a particular $|\Delta J|$. Similarly transitions within the same vibrational eigenstates have higher probabilities compared to the inter-eigenstates transitions. This is why the (dashed) curves of Fig. \ref{fig:PCL_collisions} belonging to the $\Gamma_2$ eigenstate levels are consistently higher than their (continuous) $\Gamma_1$ counterparts. The population of the initial state, $\Gamma_2, J=12$, decreases exponentially as expected but does not yet reach zero after 300 ps while the population of the $J=12$ level stays the largest one of $\Gamma_1$.
The statistical limit is almost reached within a few hundreds of ps. This means that any successful state-control targeting either the $\Gamma_1$ or $\Gamma_2$ eigenstates or any combination would therefore rapidly (\textit{i.e.} within hundreds of ps) be lost due to collision-induced transitions. 
After 300 ps, the ratio of populations $\frac{\rho_{\Gamma_1,J=12}}{\rho_{\Gamma_2,J=12}}=0.907$ although the energy of the $\Gamma_1$ eigenstates are lower than the $\Gamma_2$ ones. However, after 1500 ps, the same ratio evolved to 1.02 and would thus very slowly approach the ratio given by the Boltzmann distribution at $T=298$ K : $\frac{\rho_{\Gamma_1,J=12}}{\rho_{\Gamma_2,J=12}}=\frac{\exp(\frac{-E_{\Gamma_1,J=12}}{k_BT})}{\exp(\frac{-E_{\Gamma_2,J=12}}{k_BT})}=1.06$. This is due to the characteristic of the collision-induced dynamics enunciated before, intra-eigenstates transfers are stronger than inter-eigenstates transfers. Therefore the populations within the $\Gamma_2$ levels will remain larger given that the simulation starts from the singly populated $\Gamma_2,J=12$ state although the $\Gamma_1$ levels are lower in energy and should be more populated for a given $J$ at thermal equilibrium. Such equilibrium would not be reached within the characteristic timescale of the present dynamics and the construction of the Hamiltonian would not be adequate anymore, \textit{i.e.} inter-polyad couplings due to Coriolis effects could not be neglected anymore beyond the nanosecond.\cite{jpcrd_45_023103,mp_116_2213}

So far, the attention was centered on the eigenstate population dynamics driven by the collision-induced state-to-state relaxations. However, there is significant interest analyzing the dynamics of a coherent superposition of states, in the vibrational-rotational manifold. As an illustration, we constructed a coherent superposition of rotational levels within the state basis mentioned. Given that: 
\begin{equation}
\rho^\text{superpos}(t_0)=\sum_{i,j}c_i(t_0)c_j^*(t_0)|i\rangle\langle j|,
\label{eq:superpos}
\end{equation}
diagonal elements of the density matrix are the populations of the eigenstates and its non-diagonal elements are the coherences describing the evolution of coherent superpositions. 
The coefficients $c$ were chosen such that the initial populations of the $J=10-13$ levels in the $\Gamma_2$ eigenstate represent a coherent state generated after an excitation from the normalized ($\sum\rho_{ii}=1$) Boltzmann distribution in the  ground vibrational state at $T=298$ K and takes into account the intensity alternation due to ortho and para acetylene. The relative equilibrium populations in the ground state calculated for for ortho ($J$ odd) and para ($J$ even) are given by\cite{herman_handbook}
\begin{align}
\rho_{J_{\text{ortho}}}&=3\frac{\exp\left[-\left(E_{J_{\text{ortho}}}-E_{J=1}\right)/k_BT\right]}{\sum_{J_{\text{ortho}}}\exp\left[-\left(E_{J_{\text{ortho}}}-E_{J=1}\right)/k_BT\right]}\\
\rho_{J_{\text{para}}}&=\frac{\exp\left[-\left(E_{J_{\text{para}}}-E_{J=0}\right)/k_BT\right]}{\sum_{J_{\text{para}}}\exp\left[-\left(E_{J_{\text{para}}}-E_{J=0}\right)/k_BT\right]}
\end{align}
with $k_B$ the Boltzmann constant, and $T$ the temperature. This leads to a 3:1 alternation of populations in the ground vibrational state for odd and even $J$ levels respectively. This is shifted to a 1:3 (odd:even) alternation in our constructed coherent state in the excited vibrational state assuming that the superposition is produced by absorption of a coherent laser field and one-photon transitions satisfying the $|\Delta J|=1$ selection rules. The coherences are further calculated from the populations (diagonal elements of the density matrix) generated by the Boltzmann distribution and satisfying Eq. \ref{eq:superpos}. A similar approach has been used by Ma et al \cite{NC_10_5780} in the case of rotational relaxation of \ce{N2O} diluted in He although interaction between coherences induced by successive nonresonant short and intense laser pulses have been considered in this case.

\begin{figure}[htbp]
\centering
\includegraphics[width=\linewidth]{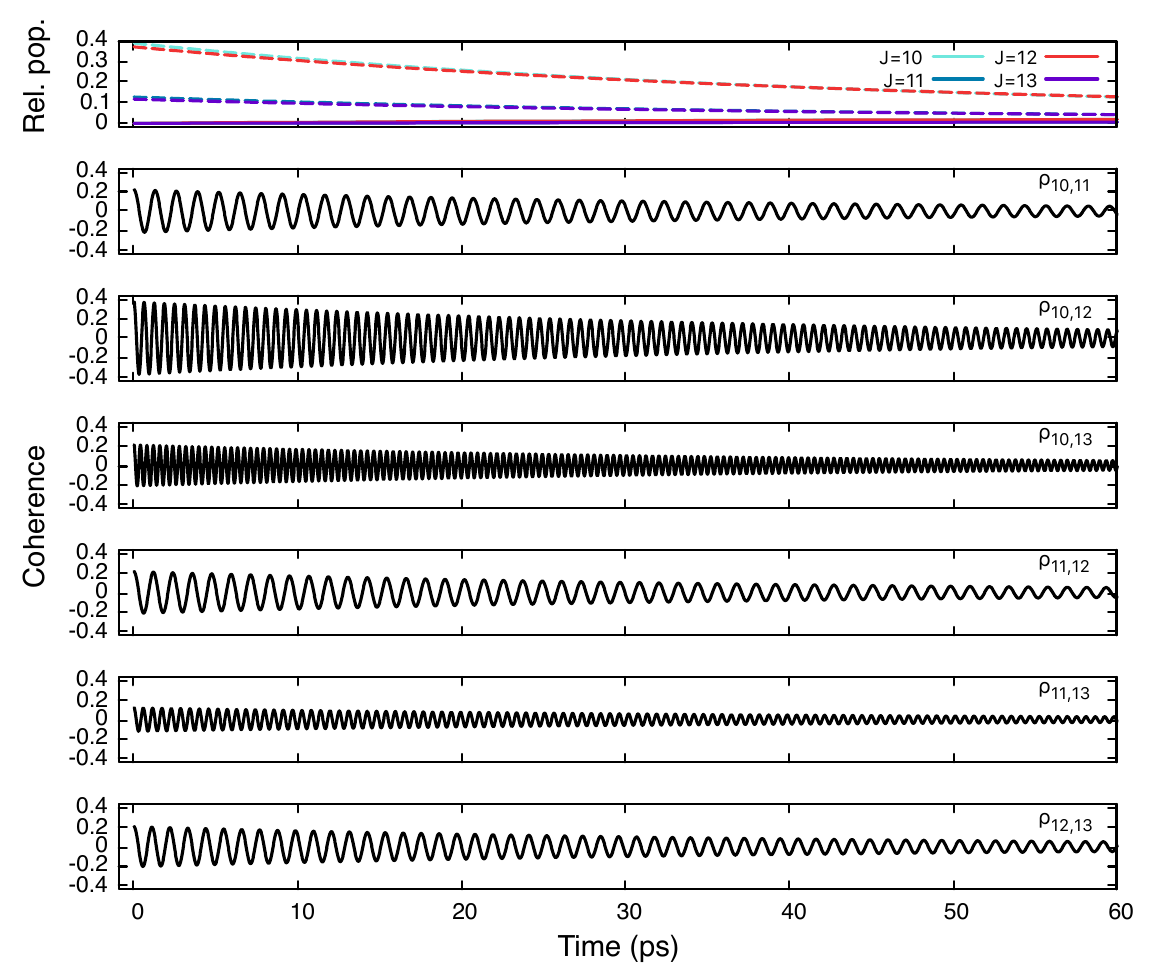} 
\caption{Collision induced dynamics in $\Gamma_1$ (\textbf{continuous}) and in $\Gamma_2$ (\textbf{dashed}) from a coherent superposition of $J=10-13$ levels in $\Gamma_2$ under $P=1$ atm and $T=298$ K. First panel shows the evolution of realtive populations within the state, later panels are the real part of coherences between the $\Gamma_2$ $J$ states initially populated (connected $J$ states are identified by subscripts). State-to-state transition rates are calculated from Eqs. \ref{equ:EGL} and \ref{EGL}, and the parameters reported by Frost \cite{jcp_98_8572} and Henton \textit{et al.}\cite{jcsft_94_3219} (reproduced in Table \ref{table:EGL_jcsft_94_3219}).}
\label{fig:decoherence}
\end{figure}

The collisions-induced dynamics is illustrated in Fig. \ref{fig:decoherence}, using the same transition rates calculated earlier, \textit{i.e.} at $P=1$ atm and $T=298$ K and the parameters of the EGL given in Table \ref{table:EGL_jcsft_94_3219}. The first panel shows the evolution of populations (diagonal elements of the density matrix) in the eigenstates. The dynamics is very similar to the one described before with the difference that it starts from a superposition of states. The following panels show the real part of the coherences between the initially populated $\Gamma_2$ levels (all others are zero by construction) identified on the figure by the corresponding $J,J'$  numbers. The time-dependent phase factor which lead to an oscillatory evolution of the coherences is given for a closed system by : $\rho_{ij}=c_i(t_0)c_j^*(t_0)\exp[-i(E_j-E_i)t]$. It is completely independent of the collision dynamics and depends only on the energy differences between the states of the basis which are the same for all the molecules of the system.

The Lindblad equation (Eq. \ref{eq:lindblad}) introduces transfers between the elements of the density matrix leading to the damping of the oscillations of coherences and hence decoherence that is expected to occur due to the interaction of the system with its environment. The off-diagonal elements of the density matrix (coherences) will follow an exponential decay to zero. The loss of coherence (decoherence) happens significantly faster than the populations relaxation in this case. Indeed, population relaxation is modulated by the population within the other states of the basis (first sum of Eq. S3 in the Supplemental Material) which implies that population flows from the most populated to the less populated states within their ``subspace" given the specific selection rules of the system. However, no such term arises from the Lindblad master equation for the evolution of coherences (Eq. S4 in the Supplemental Material). 

%-------------------------------------------------------

\section{Lindblad parameters from line broadening coefficients}
\label{section:linebroad}

As detailed in the previous section, the Lindblad parameters $\theta_{ij}$ required to model population transfers induced by collisions between molecules in the gas phase in laser and environment-driven dynamics simulations can be taken from direct kinetic measurements. However, this kind of experimental data may not be available. In this section, we describe the alternative methodology we used to determine the parameters of the EGL law, namely $K_0$ and $\eta$, from self broadening coefficients.

In this perspective, we opt for the impact and binary collision approximations, relying on the construction of the real part of the relaxation matrix and statistically based on Energy Gap fitting laws.\cite{hartmann_2008,pr_145_7} The approximations imply that the resulting perturbation on the spectrum is proportional to the gas density (binary collisions) and that the duration of the collisions is negligibly short, so that the perturbation is independent of the frequency over the spectral range considered (impact approximation). 

Using the relaxation matrix ($W$) and including the relevant factors for intensities,
the absorption coefficient $\alpha$ (in cm$^{-1}$) is given by the imaginary part of the (unnormalised) line profile:\cite{hartmann_2008,pr_145_7}
\begin{align}
\alpha(\tilde{\nu},P,T) = \ &\frac{8\pi^3}{3hc}\frac{1}{4\pi\epsilon_0}\frac{n_LT_0}{Q(T)T}[1-\exp\{-hc\tilde{\nu}/k_BT\}]\tilde{\nu} \nonumber \\
&\times \frac{1}{\pi}\sum_\ell\sum_{\ell'}\rho_\ell(T)d_\ell d_{\ell'}\left\{[\Sigma-L_a -iPW(T)]^{-1}\right\}_{\ell'\ell}
\label{eq:spectrum_intensities}
\end{align}
where $8\pi^3/(3hc)(1/(4\pi\epsilon_0))\approx4.16237\times10^{-19}$ D$^{-2}$cm$^{2}$, $n_L$ = 2.686780111$\times10^{19}$ cm$^{-3}$atm$^{-1}$ is the Loschmidt constant, $T_0=273.15$ K and $Q(T)$ is the total internal partition sum with $Q(\ce{^12C2H2}) = 412.45$ at 296 K,\cite{jcp_135_234305} $\tilde{\nu}$ is the wavenumber (in cm$^{-1}$), $P$ is the pressure of the perturber (in atm), $T$ is the temperature (in K), $\rho_\ell(T)$ is the equilibrium relative population of the initial level of line $\ell$, $d_\ell$ is the tensor that couples radiation and matter (electric dipole moment in this case, in Debye) for line $\ell$ and $i$ is the imaginary number. $\Sigma$ and $L_a$ are defined in the line space as:
\begin{align}
\Sigma_{\ell'\ell}&=\delta_{\ell,\ell'}\times \tilde{\nu} \\
\left\{L_a\right\}_{\ell'\ell}&=\delta_{\ell,\ell'}\times \tilde{\nu}_\ell
\end{align}
with $\tilde{\nu}_\ell$ the position of line $\ell$ and $\delta_{\ell,\ell'}=\delta_{i_\ell,i_{\ell'}}\phantom{.} \delta_{f_\ell,f_{\ell'}}$ where $i_{\ell}$ and $f_\ell$ are the initial and final rotational or rotation-vibration levels of the transition associated with line $\ell$. Both are diagonal matrices of size $N \times N$ in the line space, where $N$ is the number of lines.

The photon interacting with the molecules undergoing collisions is eventually dissipated in the system with a resulting broadening and shifting of the resonance wavenumber. This is modeled using the relaxation matrix $W$, which adds a complex perturbation to the resonance wavenumber within the impact approximation. The relaxation matrix is constructed with line broadening (real part) and line shift (imaginary parts) coefficients on its diagonal. The real part of the non-diagonal elements model line mixing effects on the line shape, their imaginary part being usually small and neglected.\cite{jqsrt_50_149,hartmann_2008} Only considering the diagonal elements of $W$, the spectrum calculated with Eq. \ref{eq:spectrum_intensities} is a sum of Lorentzian. The term $[\Sigma-L_a -iPW(T)]$ is a complex matrix of size $N \times N$ ($N$ is the number of lines), which must be inverted. Details on how to simulate a spectrum using Eq. \ref{eq:spectrum_intensities} along with an illustration for the $\nu_3$ band of $^{12}$C$_2$H$_2$ are provided in the Supplemental Material.

As demonstrated by Fano,\cite{pr_131_259} the line broadening coefficients $\gamma_{\ell}$ (at half width at half maximum, HWHM) of the (spectral) line $\ell$ are related to the state-to-state kinetic constants $k$ by:
\begin{equation}
\text{Re}[W_{\ell\ell}] = \gamma_\ell =\frac{1}{2}\left\{\sum_{i_{\ell'}\neq i_\ell}k(i_{\ell'}\leftarrow i_\ell)+\sum_{f_{\ell'}\neq f_\ell}k(f_{\ell'}\leftarrow f_\ell)\right\}.
\label{relax1}
\end{equation}
 The first sum describes collision-induced transitions between rotational levels in the lower vibrational level (for cold bands, $i_\ell$ levels belong to the ground vibrational state) and the second sum describes collision-induced transitions in the excited vibrational level(s). Both contributions are often assumed to be of similar magnitude. Again, the kinetic constants $k$ can be parametrized using a number of ``gap'' laws. One of the simplest is the Exponential Gap Law (EGL) (Eq. \ref{EGL}). Alternatively, a third parameter can be introduced using the Exponential Power Gap Law (EPGL) given by:
\begin{equation}
k(j\leftarrow i|K_0,\eta,\beta)=K_0\phantom{-}\left(\frac{|E_{j}-E_{i}|}{k_BT}\right)^{-\beta}\phantom{-}\exp\left(\frac{-\eta|E_{j}-E_{i}|}{k_BT}\right) \phantom{ - - } \text{with } E_j>E_i
\label{EPGL}
\end{equation}
where $K_0$ (same units as $k$), $\beta$ and $\eta$ are the fitted parameters and $E_i$ is the energy of level $i$. The downwards transition kinetic constants are deduced from the detailed balance (Eq. \ref{equ:balance}).

The interacting system $\Gamma_1 - \Gamma_2$ considered here requires some care as it may not be as straightforward to calculate the broadening coefficients from the state-to-state kinetic constants using Eq. \ref{relax1} as it may seem. The last term of the right part of Eq. \ref{relax1} must indeed include the relevant transfer channels in the excited vibrational levels described by reactions R1 to R4, \textit{e.g.} R1 and R3 for the $\Gamma_2$ eigenstate. In addition, the selection rule that forbids parity change of $J$ in collision-induced transitions must be taken into account:
\begin{eqnarray}
\label{expansion}
\sum_{f_{\ell'}\neq f_\ell}k(f_{\ell'}\leftarrow f_\ell)&=&\sum_{c=1}^{N_c}\sum_{f_{\ell'}\neq f_\ell}k_c(f_{\ell'}\leftarrow f_\ell) \\
\Delta J &=& J_{f_{\ell'}}-J_{f_{\ell}} = \text{even}\nonumber
\end{eqnarray}
where $k_c$ is the state-to-state kinetic constant within channel $c$ and $N_c$ is the number of channels ($N_c=2$ for each vibrational eigenstate here; see Eqs. \ref{equ:gamma_1} and \ref{equ:gamma_2} for the explicit sums). 

\begin{figure}[htbp]
\begin{center}
\includegraphics[width=1.0\textwidth]{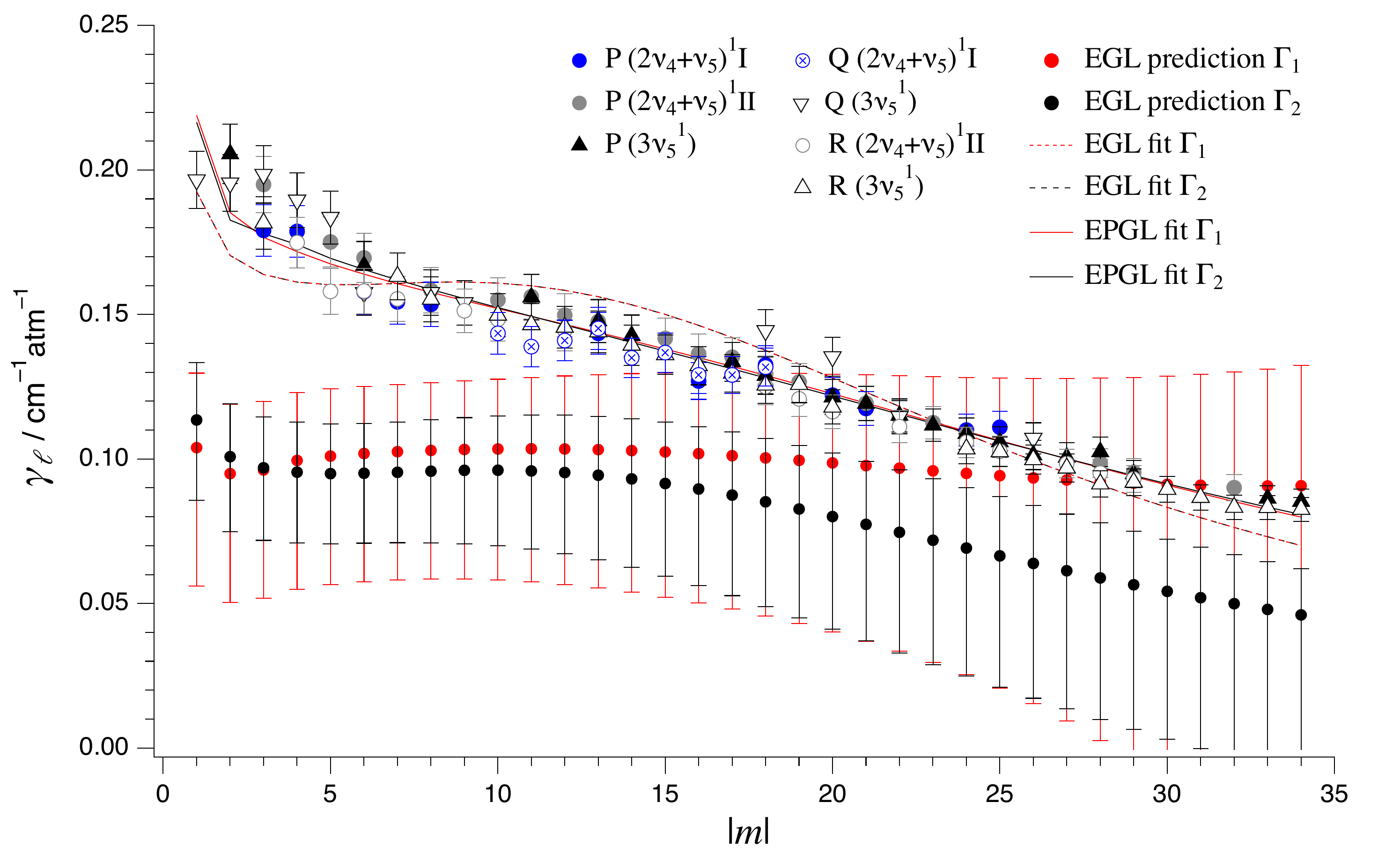}
\caption{Dependence of line broadening coefficients $\gamma_\ell$ (HWHM) with the absolute value of the rotational quantum number ($|m| = J_{i_{l}}+1$ for R branch lines and $|m| = J_{i_{l}}$ for P and Q branch lines). The experimental data measured for cold bands in the 5 $\mu$m region of acetylene\cite{jqsrt_75_397} are compared with values predicted and fitted using the EGL and EPGL (see text for details). The lines are guides for the eyes.}
\label{fig:broadexpandfit}
\end{center}
\end{figure}

Broadening coefficients have never been directly measured for the $\nu_3$ band of acetylene, most probably because of the large transition dipole moment associated with this band. However, vibrational dependence of the broadening coefficients (HWHM) is generally small, especially in acetylene.\cite{jqsrt_75_397} Therefore, we used the extensive set of experimental broadening coefficients measured for cold bands in the 5 $\mu$m region of acetylene.\cite{jqsrt_75_397} They are presented in Fig. \ref{fig:broadexpandfit}. The error bars associated with these data correspond to the upper limit of their reported precision of measurement,\cite{jqsrt_76_237} \textit{i.e.} 5 \%.

The broadening coefficients predicted using the EGL parameters of Frost\cite{jcp_98_8572} and Henton \textit{et al.}\cite{jcsft_94_3219} (see Table \ref{table:EGL_jcsft_94_3219}) are also presented in Fig. \ref{fig:broadexpandfit}. They were calculated using Eq. \ref{relax1} restricted to non radiative transitions within the excited vibrational level, \textit{i.e.}
\begin{equation}
\gamma_\ell =\sum_{f_{\ell'}\neq f_\ell}k(f_{\ell'}\leftarrow f_\ell),
\label{equ:excited_gamma}
\end{equation}
due to the lack of measurements reported for the ground vibrational state, together with Eqs. \ref{EGL}, \ref{equ:balance} and \ref{expansion}. The labels $\Gamma_1$ and $\Gamma_2$ in Fig. \ref{fig:broadexpandfit} identify the nature of the excited vibrational level. The energies of the vibrational eigenstates of interest were calculated using the effective Hamiltonian,\cite{jpcrd_45_023103}  with rotational quantum number up to $J=100$. Energies agree with reported experimental line positions\cite{jqsrt_203_3,jms_157_337} with a root mean square deviation of 0.0007 cm$^{-1}$.

Fig. \ref{fig:broadexpandfit} shows that the predictions underestimate the broadening coefficients of most lines suggesting that elastic collisions, not considered by Frost\cite{jcp_98_8572} and Henton \textit{et al.},\cite{jcsft_94_3219} contribute significantly to line broadening. Additionally, the predicted rotational dependence does not match the observation, probably because it involves extrapolation of an undersized set of measurements (only from one $J$ to others).

In view of these disagreements, the values of the parameters $K_0$ and $\eta$ involved in the EGL modeling of the transition kinetic constants $k(j \leftarrow i)$ were determined by fitting Eqs. \ref{EGL} and \ref{equ:balance} to  \ref{equ:excited_gamma} to the measured broadening coefficients\cite{jqsrt_75_397} presented in Fig. \ref{fig:broadexpandfit}. Additionally, Eq. \ref{expansion} was adapted in two ways. Orr suggested\cite{cp_190_261} that the contributions of processes such as {R}1 and {R}2 should correspond to at most 10\% of the total transition frequency from one rotational level. In Eq. \ref{expansion}, the sum of state-to-state transition frequencies of processes R1 and R2 was therefore constrained to contribute to 10\% of the broadening coefficient (processes R3 and R4 thus contribute to 90\%). Although the contributions of R1 and R2 were later refined to approximately 21\% in the polyad of interest using direct measurements of state-to-state kinetic constants,\cite{jcsft_94_3207,jcsft_94_3219} confirming Orr's suggestion that the intramolecular couplings enhance vibration -- vibration transfers,\cite{cp_190_261} significant uncertainties remain. To highlight the ability of the present methodology to describe the environment-induced processes in generic problems, this work relied on as few specific parameters as possible; this refined measurement was therefore ignored. Additionally, to keep the problem tractable and avoid strong correlations between parameters, the number of fitted parameters was restricted to two sets identified by the upper level of the transition belonging to either the $\Gamma_1$ (R2 and R4) or $\Gamma_2$ (R1 and R3) eigenstate. Taking these two constraints into account, Eqs. \ref{equ:excited_gamma} and \ref{expansion} become in the EGL:
\begin{eqnarray}
\gamma^{\Gamma_1}_\ell &=& 0.1\sum_{f_{\ell'}\neq f_\ell}k_{R2}(f^{\Gamma_2}_{\ell'}\leftarrow f^{\Gamma_1}_\ell| K_0^{\Gamma_1},\eta^{\Gamma_1}) + 0.9 \sum_{f_{\ell'}\neq f_\ell}k_{R4}(f^{\Gamma_1}_{\ell'}\leftarrow f^{\Gamma_1}_\ell|K_0^{\Gamma_1},\eta^{\Gamma_1}) \label{equ:gamma_1} \\
\gamma^{\Gamma_2}_\ell &=& 0.1 \sum_{f_{\ell'}\neq f_\ell}k_{R1}(f^{\Gamma_1}_{\ell'}\leftarrow f^{\Gamma_2}_\ell|K_0^{\Gamma_2},\eta^{\Gamma_2}) + 0.9 \sum_{f_{\ell'}\neq f_\ell}k_{R3}(f^{\Gamma_2}_{\ell'}\leftarrow f^{\Gamma_2}_\ell|K_0^{\Gamma_2},\eta^{\Gamma_2}) \label{equ:gamma_2}
\end{eqnarray}
where the transition kinetic constants are given by Eq. \ref{EGL}.
\begin{table}[h!]
\begin{center}
\begin{tabular}{cccc}
\hline
\hline
\multirow{2}{*}{}&{$K_0$}&$\eta$&$\beta$\\
\hline
EGL&  &&\\
\hline
$\Gamma_1$& 0.064(2) &1.98(5)&/\\
$\Gamma_2$& 0.064(2) &1.98(5)&/\\
\hline
EPGL&&&\\
\hline
$\Gamma_1$& 0.034(2) &1.43(5)&0.35(3)\\
$\Gamma_2$& 0.032(2) &1.37(5)&0.37(3)\\
\hline
\hline
\end{tabular}
\caption{Best-fit values of the parameters of the two Energy Gap fitting laws used in this work (see text for details). $K_0$ are in cm$^{-1}$atm$^{-1}$; the other parameters are unitless. $\Gamma_1$ and $\Gamma_2$ identify the eigenstates of the upper level of the transition (appearing as an exponent in Eqs. \ref{equ:gamma_1} and \ref{equ:gamma_2}). The numbers between parentheses are the standard deviations in the units of the last digit quoted.}
\label{table:ourfits}
\end{center}
\end{table}
All broadening coefficients reported by Jacquemart \textit{et al.}\cite{jqsrt_75_397} for cold bands in the 5 $\mu$m region of acetylene were included in the fitting procedure. The coefficients reported for the same value of $|m|$ were averaged and assigned to transitions independently of their upper level belonging to the $\Gamma_1$ or $\Gamma_2$ eigenstate. The fitting was performed using the optimization library of SciPy\cite{scipy} (scipy.optimize) with all tolerance parameters set to machine epsilon. The best-fit values of the EGL parameters obtained are given in Table \ref{table:ourfits} and the corresponding rotational dependences of the broadening coefficients are presented in Fig. \ref{fig:broadexpandfit} (dashed lines identified by ``EGL fit $\Gamma_1$'' and  ``EGL fit $\Gamma_2$,'' actually overlapped). Fig. \ref{fig:broadexpandfit} shows that, although the EGL reproduces the general decrease of the broadening coefficients with increasing rotation, it exhibits additional oscillations. It is worth to point out that the EGL cannot adequately reflect the effects of the energy differences as all parameters are the same except for the proportionality weights of 10 and 90 \% imposed to the contributions of processes R1 and R3 (for $\Gamma_2$) and R2 and R4 (for $\Gamma_1$) (Eqs. \ref{equ:gamma_1} and \ref{equ:gamma_2}). As the EGL parameters of Frost\cite{jcp_98_8572} and Henton \textit{et al.}\cite{jcsft_94_3219} yield broadening coefficients smaller than observed, it is not surprising that the present fit results in larger values of the $K_0$ and $\eta$ parameters. The fitted exponential parameters ($\eta$) however lie within the uncertainties of most reported ones (see Table \ref{table:EGL_jcsft_94_3219}).

As shown in Fig. \ref{fig:broadexpandfit}, the introduction of an additional parameter $\beta$ in the EPGL (Eq. \ref{EPGL}) results in a better description of the rotational dependence of the broadening coefficients. The resulting best-fit values of the parameters of the EPGL are also presented in Table \ref{table:ourfits}.

\begin{figure}[htbp] 
\centering
\includegraphics[width=1.1\linewidth]{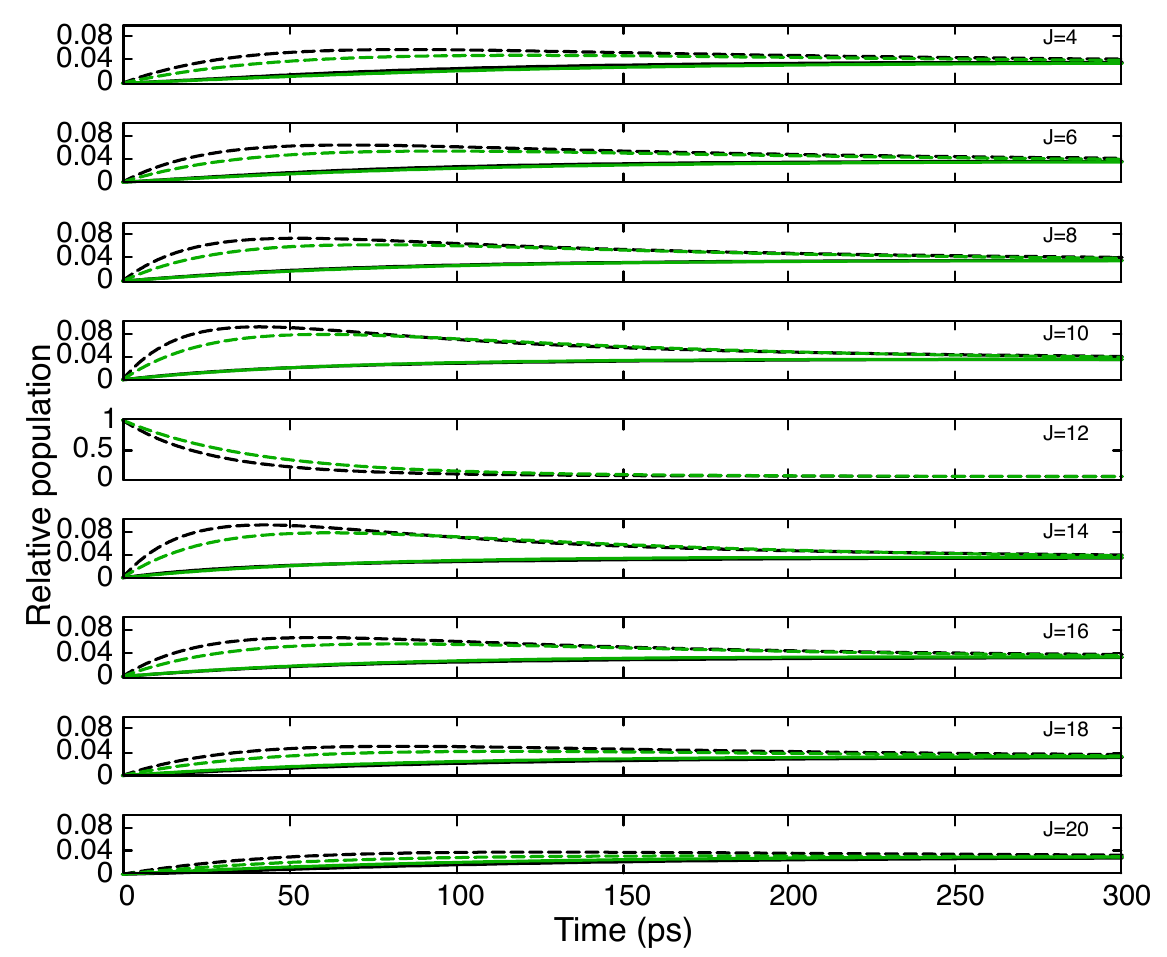}
\caption{Collision induced dynamics in $\Gamma_1$ (\textbf{continuous}) and in $\Gamma_2$ (\textbf{dashed}) from unitary population in the $|\Gamma_2, J=12\rangle$ state under $P=1$ atm. The state-to-state transition frequencies are calculated using the EPGL parameters listed in Table \ref{table:ourfits} (black) and the EGL experimental parameters reproduced in Table \ref{table:EGL_jcsft_94_3219} (green).}
\label{fig:PCL_collisions_fit}
\end{figure}

Fig. \ref{fig:PCL_collisions_fit} shows a simulation of population dynamics for selected states in the polyad of interest, similar to those presented in Fig. \ref{fig:PCL_collisions}. The Lindblad parameters are calculated using both the EPGL approach and its parameters listed in Table \ref{table:ourfits} and the EGL with its parameters listed in Table \ref{table:EGL_jcsft_94_3219}. The state basis used for the simulation is the same as for Fig. \ref{fig:PCL_collisions}. Note that our fitting procedure within the impact and binary collision approximations includes level energies up to $J$=100 and broadening coefficients up to $|m|=34$ while IRUVDR measurements\cite{jcp_98_8572,jcsft_94_3219} were fitted to the EGL model relying on transitions from a singly populated state (see Table \ref{table:EGL_jcsft_94_3219}) and including transitions with $-8\le \Delta J\le+10$ at most.

The population transfers based on the present EPGL fit compare well with those predicted using the EGL experimental model. As expected from the broadening coefficients predicted from the measurements of Frost\cite{jcp_98_8572} and Henton \textit{et al.}\cite{jcsft_94_3219} being smaller than the fitted broadening coefficients presented in Fig. \ref{fig:broadexpandfit}, the transition frequencies calculated using the EPGL model are larger than the measured transitions frequencies.\cite{jcp_98_8572,jcsft_94_3219} However, the dynamics happen on a similar timescale and the trends captured by the experiment are reproduced by the present model, \textit{i.e.} the $|\Delta J|=2$ propensity shown by the $J=10$ and $J=14$ curves that reflects the vicinity of levels energies and the faster fall of upwards transfers ($\Delta J>0$) rather than the downwards transfers ($\Delta J<0$). For the sake of comparison, after 300 ps, the ratio of populations $\frac{\rho_{\Gamma_1,J=12}}{\rho_{\Gamma_2,J=12}}=0.910$ very close to the result obtained previously.

In spite of the limitations mentioned here above, this methodology leads to good results given its simplicity, the number of parameters of the chosen fitting laws and the experimental data used as the dynamics presented are very similar.

%-------------------------------------------------------

\section{Conclusions and perspectives}
Two different methods allowing to extract the phenomenological parameters of the Lindblad equations from experimental data have been presented. In the first approach, the transitions frequencies are taken directly from a pump-probe experiment. In the second case, they are extracted from measured broadening coefficients. Both approaches rely on the use of the very simple Energy Gap laws and provide similar results with an unbeatable, inexpensive computational effort. Alternatively, the parameters could for example be calculated \textit{ab-initio}. However, this would require the computation of an interaction potential, which can be very expensive due to the number of degrees of freedom, followed by the calculation of the $S$ matrix.

The parameters obtained in this way allow to include rotational relaxation in dynamics simulations and in particular for laser control optimisations. We point out that the model being as good as its underlying approximations, it still depends strongly on assumptions. Main limitations follow that transient effects of collisions (non-Markovian effects\cite{pr_131_259}) are not included limiting the validity to a finite range of frequencies\cite{pr_145_7} and Doppler broadening is neglected so that the model does not hold at low pressure. 

In the case of the selected polyad of acetylene, we show that the rotational relaxation will induce a reorganisation of the populations initially confined in one $J$ level in about 200 ps.  Previous works on laser control of rovibrational population in acetylene\cite{mp_113_4000,mp_116_2213} have targeted the ``dark" $|0101^11^{1}\rangle$ mode of C$_2$H$_2$. 
However, this ``dark" state is not affected by the collisions on the same timescale as the states studied here due to selection rules. Indeed, the only transitions affecting this states are parity-violation transitions which happen at the second to kilosecond timescale.\cite{herman_handbook} The  $|0101^11^{1}\rangle$ mode can therefore be seen as a ``decoherence free'' subspace, which could makes it an interesting target for possible state-control studies.

%-------------------------------------------------------
\section*{Supplementary Material}
See the supplementary material for comments on the implementation of the Lindblad master equation and an illustration of the use of the relaxation matrix for the simulation of spectra.

\begin{acknowledgments}
The authors warmly thank Michel Herman for very fruitful discussions. The IISN (Institut Interuniversitaire des Sciences Nucl\'eaires) is acknowledged for its financial support. Computational resources have been provided by the Shared ICT Services Centre, Universit\'e libre de Bruxelles. %{\color{red}A.A. acknowledges the financial support of the Lucia DeBrouckere fund.} 
\end{acknowledgments}

\section*{Data Availability Statement}

The data that support the findings of this study are available from the corresponding author upon reasonable request.

%-------------------------------------------------------

%\bibliography{decoherence_lindblad.bib}

\begin{thebibliography}{55}%
\makeatletter
\providecommand \@ifxundefined [1]{%
 \@ifx{#1\undefined}
}%
\providecommand \@ifnum [1]{%
 \ifnum #1\expandafter \@firstoftwo
 \else \expandafter \@secondoftwo
 \fi
}%
\providecommand \@ifx [1]{%
 \ifx #1\expandafter \@firstoftwo
 \else \expandafter \@secondoftwo
 \fi
}%
\providecommand \natexlab [1]{#1}%
\providecommand \enquote  [1]{``#1''}%
\providecommand \bibnamefont  [1]{#1}%
\providecommand \bibfnamefont [1]{#1}%
\providecommand \citenamefont [1]{#1}%
\providecommand \href@noop [0]{\@secondoftwo}%
\providecommand \href [0]{\begingroup \@sanitize@url \@href}%
\providecommand \@href[1]{\@@startlink{#1}\@@href}%
\providecommand \@@href[1]{\endgroup#1\@@endlink}%
\providecommand \@sanitize@url [0]{\catcode `\\12\catcode `\$12\catcode
  `\&12\catcode `\#12\catcode `\^12\catcode `\_12\catcode `\%12\relax}%
\providecommand \@@startlink[1]{}%
\providecommand \@@endlink[0]{}%
\providecommand \url  [0]{\begingroup\@sanitize@url \@url }%
\providecommand \@url [1]{\endgroup\@href {#1}{\urlprefix }}%
\providecommand \urlprefix  [0]{URL }%
\providecommand \Eprint [0]{\href }%
\providecommand \doibase [0]{http://dx.doi.org/}%
\providecommand \selectlanguage [0]{\@gobble}%
\providecommand \bibinfo  [0]{\@secondoftwo}%
\providecommand \bibfield  [0]{\@secondoftwo}%
\providecommand \translation [1]{[#1]}%
\providecommand \BibitemOpen [0]{}%
\providecommand \bibitemStop [0]{}%
\providecommand \bibitemNoStop [0]{.\EOS\space}%
\providecommand \EOS [0]{\spacefactor3000\relax}%
\providecommand \BibitemShut  [1]{\csname bibitem#1\endcsname}%
\let\auto@bib@innerbib\@empty
%</preamble>
\bibitem [{\citenamefont {Koch}(2016)}]{jpcm_28_213001}%
  \BibitemOpen
  \bibfield  {author} {\bibinfo {author} {\bibfnamefont {C.~P.}\ \bibnamefont
  {Koch}},\ }\bibfield  {title} {\enquote {\bibinfo {title} {Controlling open
  quantum systems: tools, achievements, and limitations},}\ }\href@noop {}
  {\bibfield  {journal} {\bibinfo  {journal} {J. Phys. Condensed Mat.}\
  }\textbf {\bibinfo {volume} {28}},\ \bibinfo {pages} {213001} (\bibinfo
  {year} {2016})}\BibitemShut {NoStop}%
\bibitem [{\citenamefont {Koch}, \citenamefont {Lemeshko},\ and\ \citenamefont
  {Sugny}(2019)}]{rmp_91_035005}%
  \BibitemOpen
  \bibfield  {author} {\bibinfo {author} {\bibfnamefont {C.~P.}\ \bibnamefont
  {Koch}}, \bibinfo {author} {\bibfnamefont {M.}~\bibnamefont {Lemeshko}}, \
  and\ \bibinfo {author} {\bibfnamefont {D.}~\bibnamefont {Sugny}},\ }\bibfield
   {title} {\enquote {\bibinfo {title} {Quantum control of molecular
  rotation},}\ }\href@noop {} {\bibfield  {journal} {\bibinfo  {journal} {Rev.
  Mod. Phys.}\ }\textbf {\bibinfo {volume} {91}},\ \bibinfo {pages} {035005}
  (\bibinfo {year} {2019})}\BibitemShut {NoStop}%
\bibitem [{\citenamefont {Br{\"u}ggemann}, \citenamefont {Pullerits},\ and\
  \citenamefont {May}(2007)}]{jppac_190_372}%
  \BibitemOpen
  \bibfield  {author} {\bibinfo {author} {\bibfnamefont {B.}~\bibnamefont
  {Br{\"u}ggemann}}, \bibinfo {author} {\bibfnamefont {T.}~\bibnamefont
  {Pullerits}}, \ and\ \bibinfo {author} {\bibfnamefont {V.}~\bibnamefont
  {May}},\ }\bibfield  {title} {\enquote {\bibinfo {title} {Laser pulse control
  of exciton dynamics in a biological chromophore complex},}\ }\href@noop {}
  {\bibfield  {journal} {\bibinfo  {journal} {J. Photochem. Photobiol. A}\
  }\textbf {\bibinfo {volume} {190}},\ \bibinfo {pages} {372--377} (\bibinfo
  {year} {2007})}\BibitemShut {NoStop}%
\bibitem [{NJP(2014)}]{NJP_16_045007}%
  \BibitemOpen
  \bibfield  {title} {\enquote {\bibinfo {title} {Realistic and verifiable
  coherent control of excitonic states in a light-harvesting complex},}\
  }\href@noop {} {\bibfield  {journal} {\bibinfo  {journal} {New Journal
  Physics}\ }\textbf {\bibinfo {volume} {16}},\ \bibinfo {pages} {04007}
  (\bibinfo {year} {2014})}\BibitemShut {NoStop}%
\bibitem [{\citenamefont {Tremblay}\ and\ \citenamefont
  {Saalfrank}(2008)}]{pra_78_063408}%
  \BibitemOpen
  \bibfield  {author} {\bibinfo {author} {\bibfnamefont {J.~C.}\ \bibnamefont
  {Tremblay}}\ and\ \bibinfo {author} {\bibfnamefont {P.}~\bibnamefont
  {Saalfrank}},\ }\bibfield  {title} {\enquote {\bibinfo {title} {Guided
  locally optimal control of quantum dynamics in dissipative environments},}\
  }\href@noop {} {\bibfield  {journal} {\bibinfo  {journal} {Phys. Rev. A}\
  }\textbf {\bibinfo {volume} {78}},\ \bibinfo {pages} {063408} (\bibinfo
  {year} {2008})}\BibitemShut {NoStop}%
\bibitem [{\citenamefont {Hu}, \citenamefont {Gu},\ and\ \citenamefont
  {Franco}(2018)}]{jcp_148_134304}%
  \BibitemOpen
  \bibfield  {author} {\bibinfo {author} {\bibfnamefont {W.}~\bibnamefont
  {Hu}}, \bibinfo {author} {\bibfnamefont {B.}~\bibnamefont {Gu}}, \ and\
  \bibinfo {author} {\bibfnamefont {I.}~\bibnamefont {Franco}},\ }\bibfield
  {title} {\enquote {\bibinfo {title} {Lessons on electronic decoherence in
  molecules from exact modeling},}\ }\href@noop {} {\bibfield  {journal}
  {\bibinfo  {journal} {J. Chem. Phys.}\ }\textbf {\bibinfo {volume} {148}},\
  \bibinfo {pages} {134304} (\bibinfo {year} {2018})}\BibitemShut {NoStop}%
\bibitem [{\citenamefont {Tremblay}, \citenamefont {Beyvers},\ and\
  \citenamefont {Saalfrank}(2008)}]{jcp_128_194709}%
  \BibitemOpen
  \bibfield  {author} {\bibinfo {author} {\bibfnamefont {J.~C.}\ \bibnamefont
  {Tremblay}}, \bibinfo {author} {\bibfnamefont {S.}~\bibnamefont {Beyvers}}, \
  and\ \bibinfo {author} {\bibfnamefont {P.}~\bibnamefont {Saalfrank}},\
  }\bibfield  {title} {\enquote {\bibinfo {title} {{Selective excitation of
  coupled CO vibrations on a dissipative Cu(100) surface by shaped infrared
  laser pulses}},}\ }\href@noop {} {\bibfield  {journal} {\bibinfo  {journal}
  {{J. Chem. Phys.}}\ }\textbf {\bibinfo {volume} {128}},\ \bibinfo {pages}
  {194709} (\bibinfo {year} {2008})}\BibitemShut {NoStop}%
\bibitem [{\citenamefont {Scholz}\ \emph {et~al.}(2019)\citenamefont {Scholz},
  \citenamefont {Lindner}, \citenamefont {Lon\v{c}ari\'c}, \citenamefont
  {Tremblay}, \citenamefont {Juaristi}, \citenamefont {Alducin},\ and\
  \citenamefont {Saalfrank}}]{prb_100_245431}%
  \BibitemOpen
  \bibfield  {author} {\bibinfo {author} {\bibfnamefont {R.}~\bibnamefont
  {Scholz}}, \bibinfo {author} {\bibfnamefont {S.}~\bibnamefont {Lindner}},
  \bibinfo {author} {\bibfnamefont {I.}~\bibnamefont {Lon\v{c}ari\'c}},
  \bibinfo {author} {\bibfnamefont {J.-C.}\ \bibnamefont {Tremblay}}, \bibinfo
  {author} {\bibfnamefont {J.~I.}\ \bibnamefont {Juaristi}}, \bibinfo {author}
  {\bibfnamefont {M.}~\bibnamefont {Alducin}}, \ and\ \bibinfo {author}
  {\bibfnamefont {P.}~\bibnamefont {Saalfrank}},\ }\bibfield  {title} {\enquote
  {\bibinfo {title} {Vibrational response and motion of carbon monoxide on
  {Cu(100)} driven by femtosecond laser pulses: Molecular dynamics with
  electronic friction},}\ }\href@noop {} {\bibfield  {journal} {\bibinfo
  {journal} {Phys. Rev. B}\ }\textbf {\bibinfo {volume} {100}},\ \bibinfo
  {pages} {245431} (\bibinfo {year} {2019})}\BibitemShut {NoStop}%
\bibitem [{\citenamefont {Chenel}\ \emph {et~al.}(2012)\citenamefont {Chenel},
  \citenamefont {Dive}, \citenamefont {Meier},\ and\ \citenamefont
  {Desouter-Lecomte}}]{jpca_116_11273}%
  \BibitemOpen
  \bibfield  {author} {\bibinfo {author} {\bibfnamefont {A.}~\bibnamefont
  {Chenel}}, \bibinfo {author} {\bibfnamefont {G.}~\bibnamefont {Dive}},
  \bibinfo {author} {\bibfnamefont {C.}~\bibnamefont {Meier}}, \ and\ \bibinfo
  {author} {\bibfnamefont {M.}~\bibnamefont {Desouter-Lecomte}},\ }\bibfield
  {title} {\enquote {\bibinfo {title} {Control in a dissipative environment:
  The example of a {C}ope rearrangement},}\ }\href@noop {} {\bibfield
  {journal} {\bibinfo  {journal} {J. Phys. Chem. A}\ }\textbf {\bibinfo
  {volume} {116}},\ \bibinfo {pages} {11273--11282} (\bibinfo {year}
  {2012})}\BibitemShut {NoStop}%
\bibitem [{\citenamefont {Thallmair}\ \emph {et~al.}(2017)\citenamefont
  {Thallmair}, \citenamefont {Keefer}, \citenamefont {Rott},\ and\
  \citenamefont {de~Vivie-Riedle}}]{jpb_50_082001}%
  \BibitemOpen
  \bibfield  {author} {\bibinfo {author} {\bibfnamefont {S.}~\bibnamefont
  {Thallmair}}, \bibinfo {author} {\bibfnamefont {D.}~\bibnamefont {Keefer}},
  \bibinfo {author} {\bibfnamefont {F.}~\bibnamefont {Rott}}, \ and\ \bibinfo
  {author} {\bibfnamefont {F.}~\bibnamefont {de~Vivie-Riedle}},\ }\bibfield
  {title} {\enquote {\bibinfo {title} {Simulating the control of molecular
  reactions via modulated light fields: from gas phase to solution},}\
  }\href@noop {} {\bibfield  {journal} {\bibinfo  {journal} {J. Phys. B: Atom.
  Mol. Opt. Phys.}\ }\textbf {\bibinfo {volume} {50}},\ \bibinfo {pages}
  {082001} (\bibinfo {year} {2017})}\BibitemShut {NoStop}%
\bibitem [{\citenamefont {Ramakrishna}\ and\ \citenamefont
  {Seideman}(2005)}]{prl_95_113001}%
  \BibitemOpen
  \bibfield  {author} {\bibinfo {author} {\bibfnamefont {S.}~\bibnamefont
  {Ramakrishna}}\ and\ \bibinfo {author} {\bibfnamefont {T.}~\bibnamefont
  {Seideman}},\ }\bibfield  {title} {\enquote {\bibinfo {title} {Intense laser
  alignment in dissipative media as a route to solvent properties},}\
  }\href@noop {} {\bibfield  {journal} {\bibinfo  {journal} {Phys. Rev. Lett.}\
  }\textbf {\bibinfo {volume} {95}},\ \bibinfo {pages} {113001} (\bibinfo
  {year} {2005})}\BibitemShut {NoStop}%
\bibitem [{\citenamefont {Shuang}\ and\ \citenamefont
  {Rabitz}(2006)}]{jcp_124_154105}%
  \BibitemOpen
  \bibfield  {author} {\bibinfo {author} {\bibfnamefont {F.}~\bibnamefont
  {Shuang}}\ and\ \bibinfo {author} {\bibfnamefont {H.}~\bibnamefont
  {Rabitz}},\ }\bibfield  {title} {\enquote {\bibinfo {title} {Cooperating or
  fighting with decoherence in optimal control of quantum dynamics},}\
  }\href@noop {} {\bibfield  {journal} {\bibinfo  {journal} {J. Chem. Phys.}\
  }\textbf {\bibinfo {volume} {124}},\ \bibinfo {pages} {154105} (\bibinfo
  {year} {2006})}\BibitemShut {NoStop}%
\bibitem [{\citenamefont {Santos}\ \emph
  {et~al.}(2015{\natexlab{a}})\citenamefont {Santos}, \citenamefont {Justum},
  \citenamefont {Vaeck},\ and\ \citenamefont
  {Desouter-Lecomte}}]{jcp_142_134304}%
  \BibitemOpen
  \bibfield  {author} {\bibinfo {author} {\bibfnamefont {L.}~\bibnamefont
  {Santos}}, \bibinfo {author} {\bibfnamefont {Y.}~\bibnamefont {Justum}},
  \bibinfo {author} {\bibfnamefont {N.}~\bibnamefont {Vaeck}}, \ and\ \bibinfo
  {author} {\bibfnamefont {M.}~\bibnamefont {Desouter-Lecomte}},\ }\bibfield
  {title} {\enquote {\bibinfo {title} {Simulation of the elementary evolution
  operator with the motional states of an ion in an anharmonic trap},}\
  }\href@noop {} {\bibfield  {journal} {\bibinfo  {journal} {J. Chem. Phys.}\
  }\textbf {\bibinfo {volume} {142}},\ \bibinfo {pages} {134304} (\bibinfo
  {year} {2015}{\natexlab{a}})}\BibitemShut {NoStop}%
\bibitem [{\citenamefont {Santos}\ \emph
  {et~al.}(2015{\natexlab{b}})\citenamefont {Santos}, \citenamefont
  {Iacobellis}, \citenamefont {Herman}, \citenamefont {Perry}, \citenamefont
  {Desouter-Lecomte},\ and\ \citenamefont {Vaeck}}]{mp_113_4000}%
  \BibitemOpen
  \bibfield  {author} {\bibinfo {author} {\bibfnamefont {L.}~\bibnamefont
  {Santos}}, \bibinfo {author} {\bibfnamefont {N.}~\bibnamefont {Iacobellis}},
  \bibinfo {author} {\bibfnamefont {M.}~\bibnamefont {Herman}}, \bibinfo
  {author} {\bibfnamefont {D.}~\bibnamefont {Perry}}, \bibinfo {author}
  {\bibfnamefont {M.}~\bibnamefont {Desouter-Lecomte}}, \ and\ \bibinfo
  {author} {\bibfnamefont {N.}~\bibnamefont {Vaeck}},\ }\bibfield  {title}
  {\enquote {\bibinfo {title} {{A test of optimal laser impulsion for
  controlling population within the $N_s$= 1, $N_r$= 5 polyad of
  $^{12}$C$_2$H$_2$}},}\ }\href@noop {} {\bibfield  {journal} {\bibinfo
  {journal} {Mol. Phys.}\ }\textbf {\bibinfo {volume} {113}},\ \bibinfo {pages}
  {4000--4006} (\bibinfo {year} {2015}{\natexlab{b}})}\BibitemShut {NoStop}%
\bibitem [{\citenamefont {Santos}\ \emph {et~al.}(2018)\citenamefont {Santos},
  \citenamefont {Herman}, \citenamefont {Desouter-Lecomte},\ and\ \citenamefont
  {Vaeck}}]{mp_116_2213}%
  \BibitemOpen
  \bibfield  {author} {\bibinfo {author} {\bibfnamefont {L.}~\bibnamefont
  {Santos}}, \bibinfo {author} {\bibfnamefont {M.}~\bibnamefont {Herman}},
  \bibinfo {author} {\bibfnamefont {M.}~\bibnamefont {Desouter-Lecomte}}, \
  and\ \bibinfo {author} {\bibfnamefont {N.}~\bibnamefont {Vaeck}},\ }\bibfield
   {title} {\enquote {\bibinfo {title} {{Rovibrational laser control targeting
  a dark state in acetylene. Simulation in the $N_s$= 1, $N_r$= 5 polyad}},}\
  }\href@noop {} {\bibfield  {journal} {\bibinfo  {journal} {Mol. Phys.}\
  }\textbf {\bibinfo {volume} {116}},\ \bibinfo {pages} {2213--2225} (\bibinfo
  {year} {2018})}\BibitemShut {NoStop}%
\bibitem [{\citenamefont {Michelson}(1895)}]{aj_2_251}%
  \BibitemOpen
  \bibfield  {author} {\bibinfo {author} {\bibfnamefont {A.~A.}\ \bibnamefont
  {Michelson}},\ }\bibfield  {title} {\enquote {\bibinfo {title} {On the
  broadening of spectral lines},}\ }\href@noop {} {\bibfield  {journal}
  {\bibinfo  {journal} {Astrophys. J.}\ }\textbf {\bibinfo {volume} {2}},\
  \bibinfo {pages} {251} (\bibinfo {year} {1895})}\BibitemShut {NoStop}%
\bibitem [{\citenamefont {Ma}\ \emph {et~al.}(2019)\citenamefont {Ma},
  \citenamefont {Zhang}, \citenamefont {Lavorel}, \citenamefont {Billard},
  \citenamefont {Hertz}, \citenamefont {Wu}, \citenamefont {Boulet},
  \citenamefont {Hartmann},\ and\ \citenamefont {Faucher}}]{NC_10_5780}%
  \BibitemOpen
  \bibfield  {author} {\bibinfo {author} {\bibfnamefont {J.}~\bibnamefont
  {Ma}}, \bibinfo {author} {\bibfnamefont {H.}~\bibnamefont {Zhang}}, \bibinfo
  {author} {\bibfnamefont {B.}~\bibnamefont {Lavorel}}, \bibinfo {author}
  {\bibfnamefont {F.}~\bibnamefont {Billard}}, \bibinfo {author} {\bibfnamefont
  {E.}~\bibnamefont {Hertz}}, \bibinfo {author} {\bibfnamefont
  {J.}~\bibnamefont {Wu}}, \bibinfo {author} {\bibfnamefont {C.}~\bibnamefont
  {Boulet}}, \bibinfo {author} {\bibfnamefont {J.-M.}\ \bibnamefont
  {Hartmann}}, \ and\ \bibinfo {author} {\bibfnamefont {O.}~\bibnamefont
  {Faucher}},\ }\bibfield  {title} {\enquote {\bibinfo {title} {Observing
  collisions beyond the secular approximation limit},}\ }\href@noop {}
  {\bibfield  {journal} {\bibinfo  {journal} {Nat. Commun.}\ }\textbf {\bibinfo
  {volume} {10}},\ \bibinfo {pages} {5780} (\bibinfo {year}
  {2019})}\BibitemShut {NoStop}%
\bibitem [{\citenamefont {Hartmann}\ and\ \citenamefont
  {Boulet}(2012)}]{jcp_136_184302}%
  \BibitemOpen
  \bibfield  {author} {\bibinfo {author} {\bibfnamefont {J.-M.}\ \bibnamefont
  {Hartmann}}\ and\ \bibinfo {author} {\bibfnamefont {C.}~\bibnamefont
  {Boulet}},\ }\bibfield  {title} {\enquote {\bibinfo {title} {Quantum and
  classical approaches for rotational relaxation and nonresonant laser
  alignment of linear molecules: A comparison for {CO$_2$} gas in the
  nonadiabatic regime},}\ }\href@noop {} {\bibfield  {journal} {\bibinfo
  {journal} {J. Chem. Phys.}\ }\textbf {\bibinfo {volume} {136}},\ \bibinfo
  {pages} {184302} (\bibinfo {year} {2012})}\BibitemShut {NoStop}%
\bibitem [{\citenamefont {Orr}\ and\ \citenamefont {Nutt}(1980)}]{jms_84_272}%
  \BibitemOpen
  \bibfield  {author} {\bibinfo {author} {\bibfnamefont {B.~J.}\ \bibnamefont
  {Orr}}\ and\ \bibinfo {author} {\bibfnamefont {G.~F.}\ \bibnamefont {Nutt}},\
  }\bibfield  {title} {\enquote {\bibinfo {title} {Rotationally resolved
  infrared-ultraviolet double resonance spectroscopy in molecular {D$_2$CO} and
  {HDCO}},}\ }\href@noop {} {\bibfield  {journal} {\bibinfo  {journal} {J. Mol.
  Spectrosc.}\ }\textbf {\bibinfo {volume} {84}},\ \bibinfo {pages} {272--287}
  (\bibinfo {year} {1980})}\BibitemShut {NoStop}%
\bibitem [{\citenamefont {Karczmarek}\ \emph {et~al.}(1999)\citenamefont
  {Karczmarek}, \citenamefont {Wright}, \citenamefont {Corkum},\ and\
  \citenamefont {Ivanov}}]{prl_82_3420}%
  \BibitemOpen
  \bibfield  {author} {\bibinfo {author} {\bibfnamefont {J.}~\bibnamefont
  {Karczmarek}}, \bibinfo {author} {\bibfnamefont {J.}~\bibnamefont {Wright}},
  \bibinfo {author} {\bibfnamefont {P.}~\bibnamefont {Corkum}}, \ and\ \bibinfo
  {author} {\bibfnamefont {M.}~\bibnamefont {Ivanov}},\ }\bibfield  {title}
  {\enquote {\bibinfo {title} {Optical centrifuge for molecules},}\ }\href@noop
  {} {\bibfield  {journal} {\bibinfo  {journal} {Phys. Rev. Lett.}\ }\textbf
  {\bibinfo {volume} {82}},\ \bibinfo {pages} {3420} (\bibinfo {year}
  {1999})}\BibitemShut {NoStop}%
\bibitem [{\citenamefont {Villeneuve}\ \emph {et~al.}(2000)\citenamefont
  {Villeneuve}, \citenamefont {Aseyev}, \citenamefont {Dietrich}, \citenamefont
  {Spanner}, \citenamefont {Ivanov},\ and\ \citenamefont
  {Corkum}}]{prl_85_542}%
  \BibitemOpen
  \bibfield  {author} {\bibinfo {author} {\bibfnamefont {D.}~\bibnamefont
  {Villeneuve}}, \bibinfo {author} {\bibfnamefont {S.}~\bibnamefont {Aseyev}},
  \bibinfo {author} {\bibfnamefont {P.}~\bibnamefont {Dietrich}}, \bibinfo
  {author} {\bibfnamefont {M.}~\bibnamefont {Spanner}}, \bibinfo {author}
  {\bibfnamefont {M.~Y.}\ \bibnamefont {Ivanov}}, \ and\ \bibinfo {author}
  {\bibfnamefont {P.}~\bibnamefont {Corkum}},\ }\bibfield  {title} {\enquote
  {\bibinfo {title} {Forced molecular rotation in an optical centrifuge},}\
  }\href@noop {} {\bibfield  {journal} {\bibinfo  {journal} {Phys. Rev. Lett.}\
  }\textbf {\bibinfo {volume} {85}},\ \bibinfo {pages} {542} (\bibinfo {year}
  {2000})}\BibitemShut {NoStop}%
\bibitem [{\citenamefont {Shafer}\ and\ \citenamefont
  {Gordon}(1973)}]{jcp_58_5422}%
  \BibitemOpen
  \bibfield  {author} {\bibinfo {author} {\bibfnamefont {R.}~\bibnamefont
  {Shafer}}\ and\ \bibinfo {author} {\bibfnamefont {R.~G.}\ \bibnamefont
  {Gordon}},\ }\bibfield  {title} {\enquote {\bibinfo {title} {Quantum
  scattering theory of rotational relaxation and spectral line shapes in
  {H}$_2$--{He} gas mixtures},}\ }\href@noop {} {\bibfield  {journal} {\bibinfo
   {journal} {{J. Chem. Phys.}}\ }\textbf {\bibinfo {volume} {58}},\ \bibinfo
  {pages} {5422--5443} (\bibinfo {year} {1973})}\BibitemShut {NoStop}%
\bibitem [{\citenamefont {L{\'e}vy}, \citenamefont {Lacome},\ and\
  \citenamefont {Chackerian~Jr}(1992)}]{levy_1992}%
  \BibitemOpen
  \bibfield  {author} {\bibinfo {author} {\bibfnamefont {A.}~\bibnamefont
  {L{\'e}vy}}, \bibinfo {author} {\bibfnamefont {N.}~\bibnamefont {Lacome}}, \
  and\ \bibinfo {author} {\bibfnamefont {C.}~\bibnamefont {Chackerian~Jr}},\
  }\enquote {\bibinfo {title} {Collisional line mixing},}\ in\ \href@noop {}
  {\emph {\bibinfo {booktitle} {{Spectroscopy of the {E}arth's atmosphere and
  interstellar medium}}}}\ (\bibinfo  {publisher} {Academic Press Boston},\
  \bibinfo {year} {1992})\ pp.\ \bibinfo {pages} {261--337}\BibitemShut
  {NoStop}%
\bibitem [{\citenamefont {Hartmann}, \citenamefont {Boulet},\ and\
  \citenamefont {Robert}(2008)}]{hartmann_2008}%
  \BibitemOpen
  \bibfield  {author} {\bibinfo {author} {\bibfnamefont {J.-M.}\ \bibnamefont
  {Hartmann}}, \bibinfo {author} {\bibfnamefont {C.}~\bibnamefont {Boulet}}, \
  and\ \bibinfo {author} {\bibfnamefont {D.}~\bibnamefont {Robert}},\
  }\href@noop {} {\emph {\bibinfo {title} {Collisional {E}ffects on {M}olecular
  {S}pectra: {L}aboratory {E}xperiments and {M}odels, {C}onsequences for
  {A}pplications}}}\ (\bibinfo  {publisher} {Elsevier},\ \bibinfo {year}
  {2008})\BibitemShut {NoStop}%
\bibitem [{\citenamefont {Polanyi}\ and\ \citenamefont
  {Woodall}(1972)}]{jcp_56_1563}%
  \BibitemOpen
  \bibfield  {author} {\bibinfo {author} {\bibfnamefont {J.}~\bibnamefont
  {Polanyi}}\ and\ \bibinfo {author} {\bibfnamefont {K.}~\bibnamefont
  {Woodall}},\ }\bibfield  {title} {\enquote {\bibinfo {title} {Mechanism of
  rotational relaxation},}\ }\href@noop {} {\bibfield  {journal} {\bibinfo
  {journal} {J. Chem. Phys.}\ }\textbf {\bibinfo {volume} {56}},\ \bibinfo
  {pages} {1563--1572} (\bibinfo {year} {1972})}\BibitemShut {NoStop}%
\bibitem [{\citenamefont {Brunner}\ and\ \citenamefont
  {Pritchard}(1982)}]{brunner_1982}%
  \BibitemOpen
  \bibfield  {author} {\bibinfo {author} {\bibfnamefont {T.~A.}\ \bibnamefont
  {Brunner}}\ and\ \bibinfo {author} {\bibfnamefont {D.}~\bibnamefont
  {Pritchard}},\ }\enquote {\bibinfo {title} {Fitting laws for rotationally
  inelastic collisions},}\ in\ \href@noop {} {\emph {\bibinfo {booktitle}
  {Advances in Chemical Physics}}}\ (\bibinfo  {publisher} {John Wiley \& Sons,
  Ltd},\ \bibinfo {year} {1982})\ pp.\ \bibinfo {pages} {589--641}\BibitemShut
  {NoStop}%
\bibitem [{\citenamefont {Cousin}\ \emph {et~al.}(1986)\citenamefont {Cousin},
  \citenamefont {Le~Doucen}, \citenamefont {Boulet}, \citenamefont {Henry},\
  and\ \citenamefont {Robert}}]{jqsrt_36_521}%
  \BibitemOpen
  \bibfield  {author} {\bibinfo {author} {\bibfnamefont {C.}~\bibnamefont
  {Cousin}}, \bibinfo {author} {\bibfnamefont {R.}~\bibnamefont {Le~Doucen}},
  \bibinfo {author} {\bibfnamefont {C.}~\bibnamefont {Boulet}}, \bibinfo
  {author} {\bibfnamefont {A.}~\bibnamefont {Henry}}, \ and\ \bibinfo {author}
  {\bibfnamefont {D.}~\bibnamefont {Robert}},\ }\bibfield  {title} {\enquote
  {\bibinfo {title} {{Line coupling in the temperature and frequency
  dependences of absorption in the microwindows of the 4.3 $\mu$m {CO}$_2$
  band}},}\ }\href@noop {} {\bibfield  {journal} {\bibinfo  {journal} {J.
  Quant. Spectrosc. Radiat. Transf.}\ }\textbf {\bibinfo {volume} {36}},\
  \bibinfo {pages} {521--538} (\bibinfo {year} {1986})}\BibitemShut {NoStop}%
\bibitem [{\citenamefont {Strow}\ and\ \citenamefont
  {Gentry}(1986)}]{jcp_84_1149}%
  \BibitemOpen
  \bibfield  {author} {\bibinfo {author} {\bibfnamefont {L.~L.}\ \bibnamefont
  {Strow}}\ and\ \bibinfo {author} {\bibfnamefont {B.~M.}\ \bibnamefont
  {Gentry}},\ }\bibfield  {title} {\enquote {\bibinfo {title} {{Rotational
  collisional narrowing in an infrared CO$_2$ Q branch studied with a
  tunable-diode laser}},}\ }\href@noop {} {\bibfield  {journal} {\bibinfo
  {journal} {J. Chem. Phys.}\ }\textbf {\bibinfo {volume} {84}},\ \bibinfo
  {pages} {1149--1156} (\bibinfo {year} {1986})}\BibitemShut {NoStop}%
\bibitem [{\citenamefont {Gentry}\ and\ \citenamefont
  {Strow}(1987)}]{jcp_86_5722}%
  \BibitemOpen
  \bibfield  {author} {\bibinfo {author} {\bibfnamefont {B.~M.}\ \bibnamefont
  {Gentry}}\ and\ \bibinfo {author} {\bibfnamefont {L.~L.}\ \bibnamefont
  {Strow}},\ }\bibfield  {title} {\enquote {\bibinfo {title} {{Line mixing in a
  {N}$_2$-broadened {CO}$_2$ Q branch observed with a tunable diode laser}},}\
  }\href@noop {} {\bibfield  {journal} {\bibinfo  {journal} {J. Chem. Phys.}\
  }\textbf {\bibinfo {volume} {86}},\ \bibinfo {pages} {5722--5730} (\bibinfo
  {year} {1987})}\BibitemShut {NoStop}%
\bibitem [{\citenamefont {Amyay}\ \emph {et~al.}(2016)\citenamefont {Amyay},
  \citenamefont {Fayt}, \citenamefont {Herman},\ and\ \citenamefont
  {Vander~Auwera}}]{jpcrd_45_023103}%
  \BibitemOpen
  \bibfield  {author} {\bibinfo {author} {\bibfnamefont {B.}~\bibnamefont
  {Amyay}}, \bibinfo {author} {\bibfnamefont {A.}~\bibnamefont {Fayt}},
  \bibinfo {author} {\bibfnamefont {M.}~\bibnamefont {Herman}}, \ and\ \bibinfo
  {author} {\bibfnamefont {J.}~\bibnamefont {Vander~Auwera}},\ }\bibfield
  {title} {\enquote {\bibinfo {title} {{Vibration-rotation spectroscopic
  database on acetylene, $\tilde{X}^1\Sigma^+_g$ ($^{12}$C$_2$H$_2$)}},}\
  }\href@noop {} {\bibfield  {journal} {\bibinfo  {journal} {J. Phys. Chem.
  Ref. Data}\ }\textbf {\bibinfo {volume} {45}},\ \bibinfo {pages} {023103}
  (\bibinfo {year} {2016})}\BibitemShut {NoStop}%
\bibitem [{\citenamefont {Vander~Auwera}\ \emph {et~al.}(1993)\citenamefont
  {Vander~Auwera}, \citenamefont {Hurtmans}, \citenamefont {Carleer},\ and\
  \citenamefont {Herman}}]{jms_157_337}%
  \BibitemOpen
  \bibfield  {author} {\bibinfo {author} {\bibfnamefont {J.}~\bibnamefont
  {Vander~Auwera}}, \bibinfo {author} {\bibfnamefont {D.}~\bibnamefont
  {Hurtmans}}, \bibinfo {author} {\bibfnamefont {M.}~\bibnamefont {Carleer}}, \
  and\ \bibinfo {author} {\bibfnamefont {M.}~\bibnamefont {Herman}},\
  }\bibfield  {title} {\enquote {\bibinfo {title} {The $\nu_3$ fundamental in
  {C}$_2${H}$_2$},}\ }\href@noop {} {\bibfield  {journal} {\bibinfo  {journal}
  {J. Mol. Specrosc.}\ }\textbf {\bibinfo {volume} {157}},\ \bibinfo {pages}
  {337--357} (\bibinfo {year} {1993})}\BibitemShut {NoStop}%
\bibitem [{\citenamefont {Frost}(1993)}]{jcp_98_8572}%
  \BibitemOpen
  \bibfield  {author} {\bibinfo {author} {\bibfnamefont {M.~J.}\ \bibnamefont
  {Frost}},\ }\bibfield  {title} {\enquote {\bibinfo {title} {{Energy transfer
  in the $3_1$, $2_14_15_1$ Fermi-resonant states of acetylene. I. Rotational
  energy transfer}},}\ }\href@noop {} {\bibfield  {journal} {\bibinfo
  {journal} {J. Chem. Phys.}\ }\textbf {\bibinfo {volume} {98}},\ \bibinfo
  {pages} {8572--8579} (\bibinfo {year} {1993})}\BibitemShut {NoStop}%
\bibitem [{\citenamefont {Henton}\ \emph {et~al.}(1998)\citenamefont {Henton},
  \citenamefont {Islam}, \citenamefont {Gatenby},\ and\ \citenamefont
  {Smith}}]{jcsft_94_3219}%
  \BibitemOpen
  \bibfield  {author} {\bibinfo {author} {\bibfnamefont {S.}~\bibnamefont
  {Henton}}, \bibinfo {author} {\bibfnamefont {M.}~\bibnamefont {Islam}},
  \bibinfo {author} {\bibfnamefont {S.}~\bibnamefont {Gatenby}}, \ and\
  \bibinfo {author} {\bibfnamefont {I.~W.}\ \bibnamefont {Smith}},\ }\bibfield
  {title} {\enquote {\bibinfo {title} {{Rotational energy transfer and
  rotationally specific vibration--vibration intradyad transfer in collisions
  of {C}$_2${H}$_2$ $\tilde{X}$ $^1\Sigma _g^+$($3_1$/$2_1 4_1 5_1$,$ J= 10$)
  with {C}$_2${H}$_2$, Ar, He and H$_2$}},}\ }\href@noop {} {\bibfield
  {journal} {\bibinfo  {journal} {J. Chem. Soc. Far. Trans.}\ }\textbf
  {\bibinfo {volume} {94}},\ \bibinfo {pages} {3219--3228} (\bibinfo {year}
  {1998})}\BibitemShut {NoStop}%
\bibitem [{\citenamefont {Bunker}(1979)}]{bunker_1979}%
  \BibitemOpen
  \bibfield  {author} {\bibinfo {author} {\bibfnamefont {P.~R.}\ \bibnamefont
  {Bunker}},\ }\href@noop {} {\emph {\bibinfo {title} {{Molecular symmetry and
  spectroscopy}}}}\ (\bibinfo  {publisher} {Academic Press, Inc.},\ \bibinfo
  {address} {New York},\ \bibinfo {year} {1979})\BibitemShut {NoStop}%
\bibitem [{\citenamefont {Jacquemart}\ \emph {et~al.}(2002)\citenamefont
  {Jacquemart}, \citenamefont {Mandin}, \citenamefont {Dana}, \citenamefont
  {R{\'e}galia-Jarlot}, \citenamefont {Thomas},\ and\ \citenamefont {Von~der
  Heyden}}]{jqsrt_75_397}%
  \BibitemOpen
  \bibfield  {author} {\bibinfo {author} {\bibfnamefont {D.}~\bibnamefont
  {Jacquemart}}, \bibinfo {author} {\bibfnamefont {J.-Y.}\ \bibnamefont
  {Mandin}}, \bibinfo {author} {\bibfnamefont {V.}~\bibnamefont {Dana}},
  \bibinfo {author} {\bibfnamefont {L.}~\bibnamefont {R{\'e}galia-Jarlot}},
  \bibinfo {author} {\bibfnamefont {X.}~\bibnamefont {Thomas}}, \ and\ \bibinfo
  {author} {\bibfnamefont {P.}~\bibnamefont {Von~der Heyden}},\ }\bibfield
  {title} {\enquote {\bibinfo {title} {Multispectrum fitting measurements of
  line parameters for 5-$\mu$m cold bands of acetylene},}\ }\href@noop {}
  {\bibfield  {journal} {\bibinfo  {journal} {J. Quant. Spectrosc. Radiat.
  Transf.}\ }\textbf {\bibinfo {volume} {75}},\ \bibinfo {pages} {397--422}
  (\bibinfo {year} {2002})}\BibitemShut {NoStop}%
\bibitem [{\citenamefont {Pliva}(1972{\natexlab{a}})}]{jmsp_44_145}%
  \BibitemOpen
  \bibfield  {author} {\bibinfo {author} {\bibfnamefont {J.}~\bibnamefont
  {Pliva}},\ }\bibfield  {title} {\enquote {\bibinfo {title} {{Spectrum of
  acetylene in the 5-micron region}},}\ }\href@noop {} {\bibfield  {journal}
  {\bibinfo  {journal} {J. Mol. Spectrosc.}\ }\textbf {\bibinfo {volume}
  {44}},\ \bibinfo {pages} {145--164} (\bibinfo {year}
  {1972}{\natexlab{a}})}\BibitemShut {NoStop}%
\bibitem [{\citenamefont {Pliva}(1972{\natexlab{b}})}]{jmsp_44_165}%
  \BibitemOpen
  \bibfield  {author} {\bibinfo {author} {\bibfnamefont {J.}~\bibnamefont
  {Pliva}},\ }\bibfield  {title} {\enquote {\bibinfo {title} {{Molecular
  constants for the bending modes of acetylene $^{12}$C$_2$H$_2$}},}\
  }\href@noop {} {\bibfield  {journal} {\bibinfo  {journal} {J. Mol.
  Spectrosc.}\ }\textbf {\bibinfo {volume} {44}},\ \bibinfo {pages} {165--182}
  (\bibinfo {year} {1972}{\natexlab{b}})}\BibitemShut {NoStop}%
\bibitem [{\citenamefont {Herman}(2007)}]{mp_105_2217}%
  \BibitemOpen
  \bibfield  {author} {\bibinfo {author} {\bibfnamefont {M.}~\bibnamefont
  {Herman}},\ }\bibfield  {title} {\enquote {\bibinfo {title} {The acetylene
  ground state saga},}\ }\href@noop {} {\bibfield  {journal} {\bibinfo
  {journal} {Mol. Phys.}\ }\textbf {\bibinfo {volume} {105}},\ \bibinfo {pages}
  {2217--2241} (\bibinfo {year} {2007})}\BibitemShut {NoStop}%
\bibitem [{\citenamefont {Lindblad}(1976)}]{cmp_48_119}%
  \BibitemOpen
  \bibfield  {author} {\bibinfo {author} {\bibfnamefont {G.}~\bibnamefont
  {Lindblad}},\ }\bibfield  {title} {\enquote {\bibinfo {title} {On the
  generators of quantum dynamical semigroups},}\ }\href@noop {} {\bibfield
  {journal} {\bibinfo  {journal} {Commun. Math. Phys.}\ }\textbf {\bibinfo
  {volume} {48}},\ \bibinfo {pages} {119--130} (\bibinfo {year}
  {1976})}\BibitemShut {NoStop}%
\bibitem [{\citenamefont {Gorini}, \citenamefont {Kossakowski},\ and\
  \citenamefont {Sudarshan}(1976)}]{jmp_17_821}%
  \BibitemOpen
  \bibfield  {author} {\bibinfo {author} {\bibfnamefont {V.}~\bibnamefont
  {Gorini}}, \bibinfo {author} {\bibfnamefont {A.}~\bibnamefont {Kossakowski}},
  \ and\ \bibinfo {author} {\bibfnamefont {E.~C.~G.}\ \bibnamefont
  {Sudarshan}},\ }\bibfield  {title} {\enquote {\bibinfo {title} {{Completely
  positive dynamical semigroups of N-level systems}},}\ }\href@noop {}
  {\bibfield  {journal} {\bibinfo  {journal} {J. Math. Phys.}\ }\textbf
  {\bibinfo {volume} {17}},\ \bibinfo {pages} {821--825} (\bibinfo {year}
  {1976})}\BibitemShut {NoStop}%
\bibitem [{\citenamefont {Oka}(1974)}]{oka_1974}%
  \BibitemOpen
  \bibfield  {author} {\bibinfo {author} {\bibfnamefont {T.}~\bibnamefont
  {Oka}},\ }\bibfield  {title} {\enquote {\bibinfo {title} {Collision-induced
  transitions between rotational levels},}\ }in\ \href@noop {} {\emph {\bibinfo
  {booktitle} {Advances in atomic and molecular physics}}},\ Vol.~\bibinfo
  {volume} {9}\ (\bibinfo  {publisher} {Elsevier},\ \bibinfo {year} {1974})\
  pp.\ \bibinfo {pages} {127--206}\BibitemShut {NoStop}%
\bibitem [{\citenamefont {Herman}\ and\ \citenamefont
  {Li\'evin}(1982)}]{jce_59_17}%
  \BibitemOpen
  \bibfield  {author} {\bibinfo {author} {\bibfnamefont {M.}~\bibnamefont
  {Herman}}\ and\ \bibinfo {author} {\bibfnamefont {J.}~\bibnamefont
  {Li\'evin}},\ }\bibfield  {title} {\enquote {\bibinfo {title} {{Acetylene --
  From intensity alternation in spectra to ortho and para molecule}},}\
  }\href@noop {} {\bibfield  {journal} {\bibinfo  {journal} {J. Chem. Educ.}\
  }\textbf {\bibinfo {volume} {59}},\ \bibinfo {pages} {17} (\bibinfo {year}
  {1982})}\BibitemShut {NoStop}%
\bibitem [{\citenamefont {Milce}\ and\ \citenamefont
  {Orr}(1997)}]{jcp_106_3592}%
  \BibitemOpen
  \bibfield  {author} {\bibinfo {author} {\bibfnamefont {A.~P.}\ \bibnamefont
  {Milce}}\ and\ \bibinfo {author} {\bibfnamefont {B.~J.}\ \bibnamefont
  {Orr}},\ }\bibfield  {title} {\enquote {\bibinfo {title} {{The
  $\nu_{\mathrm{{CC}}}$+3$\nu_{\mathrm{{CH}}}$ rovibrational manifold of
  acetylene. {I}. {C}ollision-induced state-to-state transfer kinetics}},}\
  }\href {\doibase 10.1063/1.473466} {\bibfield  {journal} {\bibinfo  {journal}
  {J. Chem. Phys.}\ }\textbf {\bibinfo {volume} {106}},\ \bibinfo {pages}
  {3592--3606} (\bibinfo {year} {1997})}\BibitemShut {NoStop}%
\bibitem [{\citenamefont {Gordon}(1966)}]{jcp_45_1649}%
  \BibitemOpen
  \bibfield  {author} {\bibinfo {author} {\bibfnamefont {R.~G.}\ \bibnamefont
  {Gordon}},\ }\bibfield  {title} {\enquote {\bibinfo {title} {Semiclassical
  theory of spectra and relaxation in molecular gases},}\ }\href {\doibase
  10.1063/1.1727808} {\bibfield  {journal} {\bibinfo  {journal} {J. Chem.
  Phys.}\ }\textbf {\bibinfo {volume} {45}},\ \bibinfo {pages} {1649--1655}
  (\bibinfo {year} {1966})}\BibitemShut {NoStop}%
\bibitem [{\citenamefont {Meinrenken}\ \emph {et~al.}(1997)\citenamefont
  {Meinrenken}, \citenamefont {Gillespie}, \citenamefont {Macheret},
  \citenamefont {Lempert},\ and\ \citenamefont {Miles}}]{jcp_106_8299}%
  \BibitemOpen
  \bibfield  {author} {\bibinfo {author} {\bibfnamefont {C.~J.}\ \bibnamefont
  {Meinrenken}}, \bibinfo {author} {\bibfnamefont {W.~D.}\ \bibnamefont
  {Gillespie}}, \bibinfo {author} {\bibfnamefont {S.}~\bibnamefont {Macheret}},
  \bibinfo {author} {\bibfnamefont {W.~R.}\ \bibnamefont {Lempert}}, \ and\
  \bibinfo {author} {\bibfnamefont {R.~B.}\ \bibnamefont {Miles}},\ }\bibfield
  {title} {\enquote {\bibinfo {title} {Time domain modeling of spectral
  collapse in high density molecular gases},}\ }\href@noop {} {\bibfield
  {journal} {\bibinfo  {journal} {J. Chem. Phys.}\ }\textbf {\bibinfo {volume}
  {106}},\ \bibinfo {pages} {8299--8309} (\bibinfo {year} {1997})}\BibitemShut
  {NoStop}%
\bibitem [{\citenamefont {Herman}(2011)}]{herman_handbook}%
  \BibitemOpen
  \bibfield  {author} {\bibinfo {author} {\bibfnamefont {M.}~\bibnamefont
  {Herman}},\ }\bibfield  {title} {\enquote {\bibinfo {title}
  {High‐resolution infrared spectroscopy of acetylene: Theoretical background
  and research trends},}\ }in\ \href@noop {} {\emph {\bibinfo {booktitle}
  {Handbook of high-resolution spectroscopy}}},\ \bibinfo {editor} {edited by\
  \bibinfo {editor} {\bibfnamefont {M.}~\bibnamefont {Quack}}\ and\ \bibinfo
  {editor} {\bibfnamefont {F.}~\bibnamefont {Merkt}}}\ (\bibinfo  {publisher}
  {John Wiley \& Sons, Ltd.},\ \bibinfo {year} {2011})\BibitemShut {NoStop}%
\bibitem [{\citenamefont {Ben-Reuven}(1966)}]{pr_145_7}%
  \BibitemOpen
  \bibfield  {author} {\bibinfo {author} {\bibfnamefont {A.}~\bibnamefont
  {Ben-Reuven}},\ }\bibfield  {title} {\enquote {\bibinfo {title} {Impact
  broadening of microwave spectra},}\ }\href@noop {} {\bibfield  {journal}
  {\bibinfo  {journal} {Phys. Rev.}\ }\textbf {\bibinfo {volume} {145}},\
  \bibinfo {pages} {7} (\bibinfo {year} {1966})}\BibitemShut {NoStop}%
\bibitem [{\citenamefont {Amyay}, \citenamefont {Fayt},\ and\ \citenamefont
  {Herman}(2011)}]{jcp_135_234305}%
  \BibitemOpen
  \bibfield  {author} {\bibinfo {author} {\bibfnamefont {B.}~\bibnamefont
  {Amyay}}, \bibinfo {author} {\bibfnamefont {A.}~\bibnamefont {Fayt}}, \ and\
  \bibinfo {author} {\bibfnamefont {M.}~\bibnamefont {Herman}},\ }\bibfield
  {title} {\enquote {\bibinfo {title} {{Accurate partition function for
  acetylene, $^{12}$C$_2$H$_2$, and related thermodynamical quantities}},}\
  }\href@noop {} {\bibfield  {journal} {\bibinfo  {journal} {J. Chem. Phys.}\
  }\textbf {\bibinfo {volume} {135}},\ \bibinfo {pages} {234305} (\bibinfo
  {year} {2011})}\BibitemShut {NoStop}%
\bibitem [{\citenamefont {Pine}(1993)}]{jqsrt_50_149}%
  \BibitemOpen
  \bibfield  {author} {\bibinfo {author} {\bibfnamefont {A.}~\bibnamefont
  {Pine}},\ }\bibfield  {title} {\enquote {\bibinfo {title} {{Self-, N$_2$-and
  Ar-broadening and line mixing in HCN and C$_2$H$_2$}},}\ }\href@noop {}
  {\bibfield  {journal} {\bibinfo  {journal} {J. Quant. Spectrosc. Radiat.
  Transf.}\ }\textbf {\bibinfo {volume} {50}},\ \bibinfo {pages} {149--166}
  (\bibinfo {year} {1993})}\BibitemShut {NoStop}%
\bibitem [{\citenamefont {Fano}(1963)}]{pr_131_259}%
  \BibitemOpen
  \bibfield  {author} {\bibinfo {author} {\bibfnamefont {U.}~\bibnamefont
  {Fano}},\ }\bibfield  {title} {\enquote {\bibinfo {title} {Pressure
  broadening as a prototype of relaxation},}\ }\href@noop {} {\bibfield
  {journal} {\bibinfo  {journal} {Phys. Rev.}\ }\textbf {\bibinfo {volume}
  {131}},\ \bibinfo {pages} {259} (\bibinfo {year} {1963})}\BibitemShut
  {NoStop}%
\bibitem [{\citenamefont {Jacquemart}\ \emph {et~al.}(2003)\citenamefont
  {Jacquemart}, \citenamefont {Mandin}, \citenamefont {Dana}, \citenamefont
  {R{\'e}galia-Jarlot}, \citenamefont {Plateaux}, \citenamefont
  {D{\'e}catoire},\ and\ \citenamefont {Rothman}}]{jqsrt_76_237}%
  \BibitemOpen
  \bibfield  {author} {\bibinfo {author} {\bibfnamefont {D.}~\bibnamefont
  {Jacquemart}}, \bibinfo {author} {\bibfnamefont {J.-Y.}\ \bibnamefont
  {Mandin}}, \bibinfo {author} {\bibfnamefont {V.}~\bibnamefont {Dana}},
  \bibinfo {author} {\bibfnamefont {L.}~\bibnamefont {R{\'e}galia-Jarlot}},
  \bibinfo {author} {\bibfnamefont {J.-J.}\ \bibnamefont {Plateaux}}, \bibinfo
  {author} {\bibfnamefont {D.}~\bibnamefont {D{\'e}catoire}}, \ and\ \bibinfo
  {author} {\bibfnamefont {L.}~\bibnamefont {Rothman}},\ }\bibfield  {title}
  {\enquote {\bibinfo {title} {The spectrum of acetylene in the 5-$\mu$m region
  from new line-parameter measurements},}\ }\href@noop {} {\bibfield  {journal}
  {\bibinfo  {journal} {J. Quant. Spectrosc. Radiat. Transf.}\ }\textbf
  {\bibinfo {volume} {76}},\ \bibinfo {pages} {237--267} (\bibinfo {year}
  {2003})}\BibitemShut {NoStop}%
\bibitem [{\citenamefont {Gordon}\ \emph {et~al.}(2017)\citenamefont {Gordon},
  \citenamefont {Rothman}, \citenamefont {Hill}, \citenamefont {Kochanov},
  \citenamefont {Tan}, \citenamefont {Bernath}, \citenamefont {Birk},
  \citenamefont {Boudon}, \citenamefont {Campargue}, \citenamefont {Chance}
  \emph {et~al.}}]{jqsrt_203_3}%
  \BibitemOpen
  \bibfield  {author} {\bibinfo {author} {\bibfnamefont {I.~E.}\ \bibnamefont
  {Gordon}}, \bibinfo {author} {\bibfnamefont {L.~S.}\ \bibnamefont {Rothman}},
  \bibinfo {author} {\bibfnamefont {C.}~\bibnamefont {Hill}}, \bibinfo {author}
  {\bibfnamefont {R.~V.}\ \bibnamefont {Kochanov}}, \bibinfo {author}
  {\bibfnamefont {Y.}~\bibnamefont {Tan}}, \bibinfo {author} {\bibfnamefont
  {P.~F.}\ \bibnamefont {Bernath}}, \bibinfo {author} {\bibfnamefont
  {M.}~\bibnamefont {Birk}}, \bibinfo {author} {\bibfnamefont {V.}~\bibnamefont
  {Boudon}}, \bibinfo {author} {\bibfnamefont {A.}~\bibnamefont {Campargue}},
  \bibinfo {author} {\bibfnamefont {K.}~\bibnamefont {Chance}},  \emph
  {et~al.},\ }\bibfield  {title} {\enquote {\bibinfo {title} {The {HITRAN}2016
  molecular spectroscopic database},}\ }\href@noop {} {\bibfield  {journal}
  {\bibinfo  {journal} {J. Quant. Spectrosc. Radiat. Transf.}\ }\textbf
  {\bibinfo {volume} {203}},\ \bibinfo {pages} {3--69} (\bibinfo {year}
  {2017})}\BibitemShut {NoStop}%
\bibitem [{\citenamefont {Orr}(1995)}]{cp_190_261}%
  \BibitemOpen
  \bibfield  {author} {\bibinfo {author} {\bibfnamefont {B.}~\bibnamefont
  {Orr}},\ }\bibfield  {title} {\enquote {\bibinfo {title} {Collision-induced
  state-to-state energy transfer in perturbed rovibrational manifolds of small
  polyatomic molecules: {M}echanistic insights and observations},}\ }\href@noop
  {} {\bibfield  {journal} {\bibinfo  {journal} {Chem. Phys.}\ }\textbf
  {\bibinfo {volume} {190}},\ \bibinfo {pages} {261--278} (\bibinfo {year}
  {1995})}\BibitemShut {NoStop}%
\bibitem [{\citenamefont {Henton}, \citenamefont {Islam},\ and\ \citenamefont
  {Smith}(1998)}]{jcsft_94_3207}%
  \BibitemOpen
  \bibfield  {author} {\bibinfo {author} {\bibfnamefont {S.}~\bibnamefont
  {Henton}}, \bibinfo {author} {\bibfnamefont {M.}~\bibnamefont {Islam}}, \
  and\ \bibinfo {author} {\bibfnamefont {I.~W.}\ \bibnamefont {Smith}},\
  }\bibfield  {title} {\enquote {\bibinfo {title} {{Relaxation within and from
  the ($3_1$/$2_ 1 4 _1 5 _1$) and ($3_1 4_1$/$2_1 4_2 5_1$) Fermi dyads in
  acetylene: Vibrational energy transfer in collisions with C$_2$H$_2$, N$_2$
  and H$_2$}},}\ }\href@noop {} {\bibfield  {journal} {\bibinfo  {journal} {J.
  Chem. Soc., Far. Trans.}\ }\textbf {\bibinfo {volume} {94}},\ \bibinfo
  {pages} {3207--3217} (\bibinfo {year} {1998})}\BibitemShut {NoStop}%
\bibitem [{\citenamefont {Jones}\ \emph {et~al.}(01  )\citenamefont {Jones},
  \citenamefont {Oliphant}, \citenamefont {Peterson} \emph {et~al.}}]{scipy}%
  \BibitemOpen
  \bibfield  {author} {\bibinfo {author} {\bibfnamefont {E.}~\bibnamefont
  {Jones}}, \bibinfo {author} {\bibfnamefont {T.}~\bibnamefont {Oliphant}},
  \bibinfo {author} {\bibfnamefont {P.}~\bibnamefont {Peterson}},  \emph
  {et~al.},\ }\href {http://www.scipy.org/} {\enquote {\bibinfo {title}
  {{SciPy}: Open source scientific tools for {Python}},}\ } (\bibinfo {year}
  {2001--})\BibitemShut {NoStop}%
\end{thebibliography}
%

\end{document}